\documentclass[aps,prl,twocolumn,longbibliography]{revtex4-1}

\usepackage{bm}
\usepackage{graphicx}
\usepackage{amsmath}
\usepackage{epstopdf}
\usepackage{amssymb}
\usepackage[utf8]{inputenc}
\usepackage[T1]{fontenc}
\usepackage{color}
\usepackage{upgreek}
\usepackage{subfigure}
\usepackage[normalem]{ulem}
\usepackage[unicode=true,colorlinks=true,citecolor=blue]{hyperref}
\usepackage{placeins}
\usepackage{lipsum}
\usepackage{epsfig}
\usepackage[utf8]{inputenc}
\usepackage{tabularx}
\usepackage{ctable}
\usepackage{makecell}
\usepackage{wrapfig}
\usepackage{physics}
\usepackage[utf8]{inputenc}
\usepackage[T1]{fontenc}
\usepackage[exponent-product=\cdot]{siunitx}
\usepackage{braket}
\usepackage{nicefrac} 			
\newcommand{\nix}[1]{}

\definecolor{greenI}{rgb}{0, .4, 0}

\usepackage[version=4]{mhchem}
\usepackage{glossaries}
\setacronymstyle{long-short}
\glsdisablehyper
\usepackage{siunitx}

\newcommand{\kWcm}[1]{\SI{#1}{\kilo\watt\per\centi\meter\squared}}

\newcommand{\NanoMeter}[1]{\SI{#1}{\nano\meter}}

\newcommand{\TeraHertz}[1]{\SI{#1}{\tera\hertz}}

\newcommand{\fittingunit}[1]{\SI[per-mode=symbol]{#1}{\micro\ampere\centi\meter\squared\per\kilo\watt}}

\newacronym{lpge}{LPGE}{linear photogalvanic effect}
\newacronym{cpge}{CPGE}{circular photogalvanic effect}
\newacronym{cpde}{CPDE}{circular photon drag effect}
\newacronym{lpde}{LPDE}{linear photon drag effect}
\newacronym{blg}{BLG}{bilayer graphene}
\newacronym{hbn}{hBN}{hexagonal boron nitride}
\newacronym{pge}{PGE}{photogalvanic effect}
\newacronym{pde}{PDE}{photon drag effect}
\newacronym{cnp}{CNP}{charge neutrality point}
\newacronym{thz}{THz}{terahertz}
\newacronym{ebl}{EBL}{electron-beam lithography}

\begin{document}

\title{Nonlinear intensity dependence of ratchet currents induced by terahertz laser radiation in bilayer graphene with asymmetric periodic grating gates}

\author{E.~M{\"o}nch$^1$, S. Hubmann$^1$, I. Yahniuk$^{1,2}$, S. Schweiss$^1$, V.V. Bel'kov$^3$, L.E. Golub$^{1}$, R.~Huber$^{1}$, J. Eroms$^1$, K. Watanabe$^4$, T. Taniguchi$^5$, D.~Weiss$^1$,
and S.D. Ganichev$^{1,2}$}
\affiliation{$^1$Terahertz Center, University of Regensburg, 93040 Regensburg, Germany}
	
\affiliation{$^2$CENTERA Laboratories, Institute of High Pressure Physics, Polish Academy of Sciences PL-01-142 Warsaw, Poland}

\affiliation{$^3$Ioffe Institute, 194021 St. Petersburg, Russia}

\affiliation{$^4$Research Center for Electronic and Optical Materials, National Institute for Materials Science, 1-1 Namiki, Tsukuba 305-0044, Japan}

\affiliation{$^5$Research Center for Materials Nanoarchitectonics, National Institute for Materials Science, 1-1 Namiki, Tsukuba 305-0044, Japan}


\begin{abstract}
We report on the observation of a nonlinear intensity dependence of the terahertz radiation induced ratchet effects in bilayer graphene with asymmetric dual grating gate lateral lattices. These nonlinear ratchet currents are studied in structures of two designs with dual grating gate fabricated on top of boron nitride encapsulated bilayer graphene and beneath it. The strength and sign of the photocurrent can be controllably varied by changing the bias voltages applied to individual dual grating subgates and the back gate. The current consists of contributions insensitive to the radiation's polarization state, defined by the orientation of the radiation electric field vector with respect to the dual grating gate metal stripes, and the circular ratchet sensitive to the radiation helicity. We show that intense terahertz radiation results in a nonlinear intensity dependence caused by electron gas heating. At room temperature the ratchet current saturates at high intensities of the order of hundreds to several hundreds of kW\,cm$^{-2}$. At $T = 4$~K, the nonlinearity manifests itself at intensities that are one or two orders of magnitude lower, moreover, the photoresponse exhibits a complex dependence on the intensity, including a saturation and even a change of sign with increasing intensity. This complexity is attributed to the interplay of the Seebeck ratchet and the dynamic carrier density redistribution, which feature different intensity dependencies and a nonlinear behavior of the sample's conductivity induced by electron gas heating. The latter is demonstrated by studying the THz photoconductivity. Our study demonstrates that graphene-based asymmetric dual grating gate devices can be used as terahertz detectors at room temperature over a wide dynamic range, spanning many orders of magnitude of terahertz radiation power. Therefore, their integration together with current-driven read-out electronics is attractive for the operation with high-power pulsed sources.
\end{abstract}

\maketitle

\section{Introduction}
\label{introduction}

The terahertz (THz) frequency range is one of the frontiers of physics, holding great promise for progress in diverse fields, such as solid-state physics, biology, and astrophysics. It also has tremendous potential in high data rate wireless communications, security applications, environmental monitoring, spectroscopy of various materials, etc. The vast majority of these applications depend on the availability of THz detectors and sources. Most recently the asymmetric comb-like dual-grating-gate (DGG) graphene-based field-effect transistor structure has been suggested as a promising room temperature detector for the use in 6G- and 7G-class high-speed wireless communication systems~\cite{Tamura2022}. Direct currents (\textit{dc}) excited by THz electric fields in graphene-based asymmetric DGG structures have been intensively studied in a wide range of frequencies from hundreds of gigahertz to tens of terahertz~\cite{Olbrich2016,Ganichev2017,Fateev2017,Fateev2019,BoubangaTombet2020,DelgadoNotario2020,DelgadoNotario2022,Morozov2021,Moench2022,Moench2023}. Experiments performed in DGG formed by periodically repeated metal stripes of different widths deposited on mono- and bilayer graphene have revealed several different sources of the rectified current namely electronic ratchet effects~\cite{Olbrich2009,Olbrich2009,Ivchenko2011,Olbrich2011,Olbrich2016,Hubmann2020,Moench2022,Moench2023} and plasmonic ratchet effects~\cite{Popov2011,Watanabe2013,Popov2013,Otsuji2013,Otsuji2013b,Kurita2014,BoubangaTombet2014,Faltermeier2015,Popov2015,Olbrich2016,DelgadoNotario2020,Moench2022,DelgadoNotario2022,Tamura2022,Moench2023}, and the photothermoelectric effect~\cite{Tamura2022,Xu2009,Mueller2010,Yan2012,Echtermeyer2014,Cai2014}. It is worth noting, that the two former effects have recently been shown to be the hydrodynamic and drift-diffusion limits of a mechanism controlled by the ratio of the electron-impurity and electron-electron scattering rates. While \textit{dc} currents excited in graphene-based DGG structures by low-power THz radiation have been actively studied, measurements at high power excitation have not been performed so far. At low intensities the \textit{dc} current in DGG THz devices scales linearly with the radiation intensity $I$, at high intensities it can be driven in a nonlinear regime, which has not yet been addressed, either experimentally or theoretically. The study of the photocurrent nonlinearity is important because, on the one hand, such results provide new insights in the mechanisms of current formation and, on the other, allow to define the detector's dynamic range, an important figure of merit, and to obtain parameters controlling it.

Here we report the observation and detailed study of the nonlinear \textit{dc} current excited by intense THz radiation in DGG devices at room and liquid helium temperatures. Our findings show that in most cases the photocurrent saturates with an increase in the radiation intensity. The saturation intensity depends on the potentials applied to DGG subgates, the back gate voltage, and the temperature. It can be tuned in a wide range from fractions of kW\,cm$^{-2}$ to MW\,cm$^{-2}$. Furthermore, at low temperatures and for specific sets of voltages applied to the gates, the photocurrent shows a nonmonotonic intensity dependence, first saturating but then reversing its sign at higher intensities. Our analysis demonstrates that the observed nonlinearities are mainly caused by electron gas heating. We show that a complex intensity dependence results from the interplay of two microscopic mechanisms:
the Seebeck ratchet caused by an inhomogeneous electron heating due to the near field $\bm E(x)$ formed by the diffraction of the radiation incident on the lateral superlattice~\cite{Ivchenko2011,Olbrich2016,Faltermeier2017} and dynamic carrier-density redistribution (DCDR)~\cite{Ivchenko2011,Faltermeier2018}. Most of the experiments have been performed using conventional structures with a DGG formed by metal stripes deposited on top of graphene (top DGG). Additionally, we fabricated and studied devices of novel design with asymmetric DGG structures fabricated beneath the graphene (bottom DGG), in which the radiation directly enters the graphene layer without having to pass through the metal grid. We show that both kinds of DGG devices exhibit similar behavior, however, the top DGG structure has a higher room temperature responsivity than the bottom DGG. Thus, in the main body of the paper we focus on the results obtained for the top DGG structures, and in the Appendix~\ref{bottomDGG} we present and discuss the data for the bottom DGG.

The paper is organized as following. In Secs.~\ref{methods} and~\ref{samples} we briefly discuss the essential features of our research methodology and the top DGG sample design. The results on the THz ratchet currents excited by low-power and high-power radiation are presented in Secs.~\ref{lowpower} and~\ref{highpower}, respectively. In Sec.~\ref{photoconductivity} we show the data on the radiation  induced photoconductivity in top DGG structures. In Sec.~\ref{discussion} we discuss experimental data in view of the microscopic background. The last section summarizes the work. We complement our paper by Appendices~\ref{DCDR}-\ref{bottomDGG}
discussing the DCDR microscopic mechanism of the ratchet current (Appendix~\ref{DCDR}), and presenting a detailed description and discussion of the results obtained on bottom gate DGG structures (Appendix~\ref{bottomDGG}).


\section{Sample description}
\label{samples}


 \begin{figure}
	\centering
	\includegraphics[width=\linewidth]{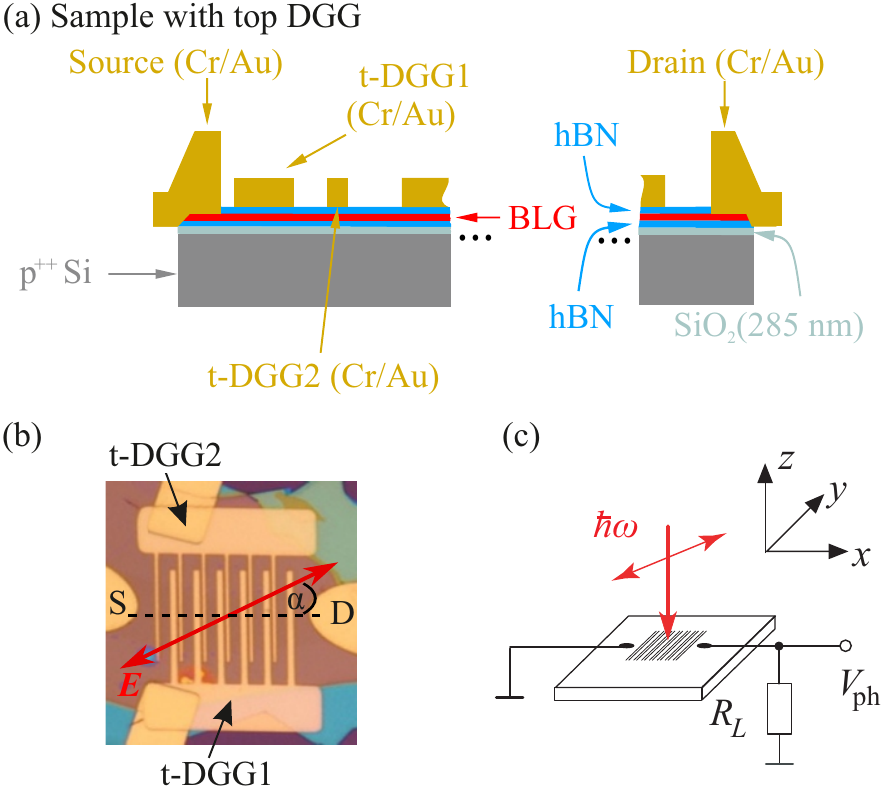}
	\caption{Panels (a) and (b) show a cross-sectional sketch and top view of the device with top DGG. The \gls{blg} structure is encapsulated in \gls{hbn} and stacked on $p$-type silicon covered by a silicon oxide layer. The dual-grating top gate structure was fabricated on top of the \gls{blg}-\gls{hbn}-sandwich and consists of Cr/Au gate fingers. The red double arrow in panel (b) depicts the orientation of the radiation electric field with respect to the source-drain (S-D) direction, which is defined by the azimuth angle $\alpha$. Panel (c) shows the measurement configuration. In the experiments with pulsed laser the voltage signal was measured across a load resistance $R_{L} = 50~\Omega$. In contrast for measurements with the $cw$ laser the voltage drop was obtained directly across the sample.}
	\label{Fig1sampletopDGG}
\end{figure}


The devices were fabricated from exfoliated bilayer graphene (BLG) flakes encapsulated in hexagonal boron nitride (hBN), which were stacked onto a \ce{Si}-substrate covered by \NanoMeter{285} of \ce{SiO_2} using the van-der-Waals stacking technique~\cite{Wang2013}. Figure~\ref{Fig1sampletopDGG} shows the cross section (a) and an optical microphotograph (b) of the BLG sandwich (the BLG size was $\approx$ 30$\times 11.5$~$\mu$m$^2$, with top and bottom hBN thicknesses of 40 and 80~nm, respectively) with the DGG on top. The Si-substrate served as a uniform back gate to control the carrier density, in the range $(0.6 - 7)\times 10^{11}$~cm$^{-2}$. The DGG is made from two electrically separated top gates, one with wide (DGG1) and the other with narrow (DGG2) stripes. The lateral lattice consists of six cells of period $L$ each consisting of wide (narrow) space and width parameters, i.e. 2~$\mu$m (0.5~$\mu$m) and 1~$\mu$m (0.5~$\mu$m), respectively. This configuration allows us to apply an asymmetric bias voltage to the sample to control the lateral asymmetry parameter $\Xi$ by using unequally biased top gates. The double comb-like structure was fabricated by using electron beam lithography (EBL) and the deposition of \NanoMeter{5} \ce{Cr} and \NanoMeter{30} \ce{Au} layers. To process the contacts, prepared by EBL, the hBN was etched by reactive ion etching and finally \ce{Au} and \ce{Cr} were deposited. 

This types of DGG have been widely used in works on THz detectors, see e.g. Refs~\cite{Watanabe2013,Faltermeier2015,DelgadoNotario2020,DelgadoNotario2022}. In addition, as addressed above, we also designed and investigated novel bottom DGG structures, see Appendix~\ref{bottomDGG} for fabrication details and experimental results.
%
The periodic grating structure allows a controllable variation of the lateral asymmetry parameter~\cite{Ivchenko2011, Olbrich2016}
\begin{equation}
\label{Xi0}
\Xi=\overline{ {dV\over dx}|\bm E_0(x)|^2},
\end{equation}
where $x$ is the axis perpendicular to the DGG stripes, $V(x)$ is the electrostatic potential and $\bm E_0(x)$ is the modulated electric field due to near-field effects caused by the diffraction of the THz radiation on the grating. In addition to the dual-grating gates, the graphene parameters were varied by a voltage applied to the back gate. 
Note that, in graphene-based structures the charge neutrality point (CNP) may shift due to residual doping within the BLG, therefore, the applied back gate voltage will be presented as an effective voltage with $U_\mathrm{BG, eff} = U_\mathrm{BG} - U^{\mathrm{max}}_{\mathrm{BG}}$ in the following, where $U^{\mathrm{max}}_{\mathrm{BG}}$ is the back gate voltage value corresponding to the CNP.

\section{Methods}
\label{methods}

The THz induced \textit{dc} current in our devices has been obtained by using normally incident low power and high power radiation from optically pumped continuous wave ($cw$)- and pulsed molecular lasers~\cite{Ganichev2005}. These line-tunable systems allowed us to generate frequency lines between 0.6 and 3.3~THz, with corresponding photon energies between 2.5 and 13.8~meV. Note that the associated wavelengths are about two orders of magnitude larger than the period of the DGG structures; a condition which is characteristic for the ratchet regime. The beam had a Gaussian shape, which was monitored by a pyroelectric camera, and was focused onto the graphene sample using an off-axis parabolic mirror resulting in a spot size of $\approx$ 1.5 to 2~mm FWHM at the sample's position. Note that the spot is much larger than the sample size ensuring approximately uniform illumination of the device. The intensities of these laser systems differ by several orders of magnitude. While for $cw$ the intensities reach $\approx$ W\,cm$^{-2}$ in the pulsed regime the peak intensities of 100~ns pulses reach a few MW\,cm$^{-2}$. The $cw$ laser with methanol as the active medium operated at $f = 2.54$~THz. We used a pulsed gas laser with NH$_3$ and CH$_3$F gases as active media pumped by a transversally excited atmospheric pressure (TEA) CO$_2$ laser~\cite{Ganichev1993,Ganichev1995,Ganichev1998}. It operates at frequencies $f = 0.6$, 1, 2, and 3.3~THz with a repetition rate of 1~Hz. Note that, for these laser lines the achievable intensity maximum decreases with frequency. The \textit{cw}-laser operates at $f=2.54$~THz and the radiation was modulated by a chopper. The power of the $cw$ laser was measured by a pyroelectric detector and that of the pulsed laser by a photon drag detector~\cite{Ganichev1985}. The initially linearly polarized radiation, was modified by a half- and quarter-wave plates, both made of $x$-cut quartz to achieve a controllable rotation of the linear polarization state and to provide elliptically/circularly polarized radiation. The radiation intensity was varied by using a crossed polarizer setup consisting of two wire grid polarizers, where the rotation of the first one modifies the radiation intensity, while the second one is fixed to ensure an unchanged output polarization~\cite{Hubmann2019, Candussio2021a}.

The samples were mounted in an optical temperature-regulated continuous flow cryostat with $z$-cut crystal quartz windows or on a room temperature sample holder. The source (S) and drain (D) contacts were arranged in such a way, that the photocurrent normal to the dual grating gate stripes could be detected, see Figs.~\ref{Fig1sampletopDGG}(a) and (b). For most of the experiments discussed below we measured the photocurrent in unbiased samples, for the setup see Fig.~\ref{Fig1sampletopDGG}(c). Additionally, we also examined the photoinduced change of the sample's conductivity. In the $cw$ laser setup the photovoltage signals $V_\text{ph}$ were measured by using a standard lock-in technique, while modulating the incoming radiation with an optical chopper. This resulting signal $V_\text{ph}$ is related to the photocurrent as $J = V_\text{ph} / R_\mathrm{s}$, where $R_{\rm s}$ is the sample resistance. In case of pulsed excitation the photocurrents were measured as a voltage drop across a load resistor $R_\mathrm{L} = 50~\Omega$, and the photocurrent is obtained as $J = V_\text{ph} / R_\mathrm{L}$. For the photoconductivity measurements the samples were additionally biased by a \textit{dc} bias voltage of $U_{\text{dc}}\pm \SI{0.03}{\volt}$. Consequently, in this case the measured signal consists of a photocurrent and a photoconductivity contributions. While the photocurrent signal is independent of $U_{\text{dc}}$, the photoconductivity signal is proportional to the bias voltage. This allowed us to extract the photoconductivity signal by taking one half of the signals obtained for negative and positive biases, which eliminates the photocurrent contribution.

\section{Photocurrents at low-power excitation}
\label{lowpower}

\begin{figure}
	\centering
	\includegraphics[width=\linewidth]{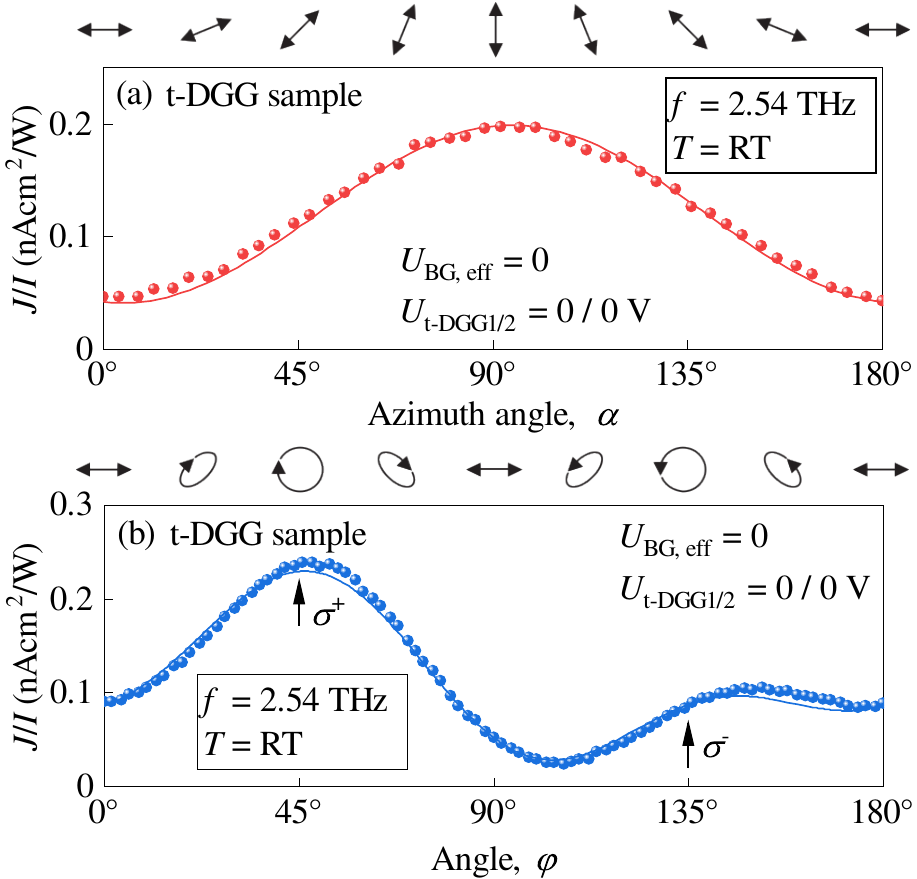}
	\caption{Photocurrent $J$ normalized to the radiation intensity $I$ measured at room temperature at a frequency of $f = 2.54$~THz. Panel (a): ratchet current as a function of the azimuth angle $\alpha$ obtained at zero effective back gate voltage corresponding to the charge neutrality point, and both top subgates unbiased. The solid line is fitted according to Eq.~\ref{linear}.  It demonstrates that the ratchet current excited by linearly polarized radiation consists of two contributions $J_{0} = 1.2$~nA\,cm$^{2}$/W and $J_{1} = 0.78$~nA\,cm$^{2}$/W. Note that a possible contribution $J_2$ is negligible. Double arrows on the top illustrate the state of polarization for several values of azimuth angles $\alpha$. Panel (b) shows the polarization dependence of the photocurrent obtained by rotating of the lambda-quarter plate by the angle $\varphi$. The solid line is fitted according to Eq.~\ref{circular}. It shows a large contribution of the circular ratchet current $J_{\rm circ} = 0.73$~nA\,cm$^{2}$/W.
Double arrows, ellipses and circles on top illustrate the state of polarization for several values of angle $\varphi$.	Upper arrows labeled as $\sigma^+$ and $\sigma^-$ indicate the photocurrent magnitude measured for right- and left-handed circularly polarized radiation, respectively.
}
	\label{lin_pol_dep}
\end{figure}

\begin{figure}[t]
	\centering
	\includegraphics[width=\linewidth]{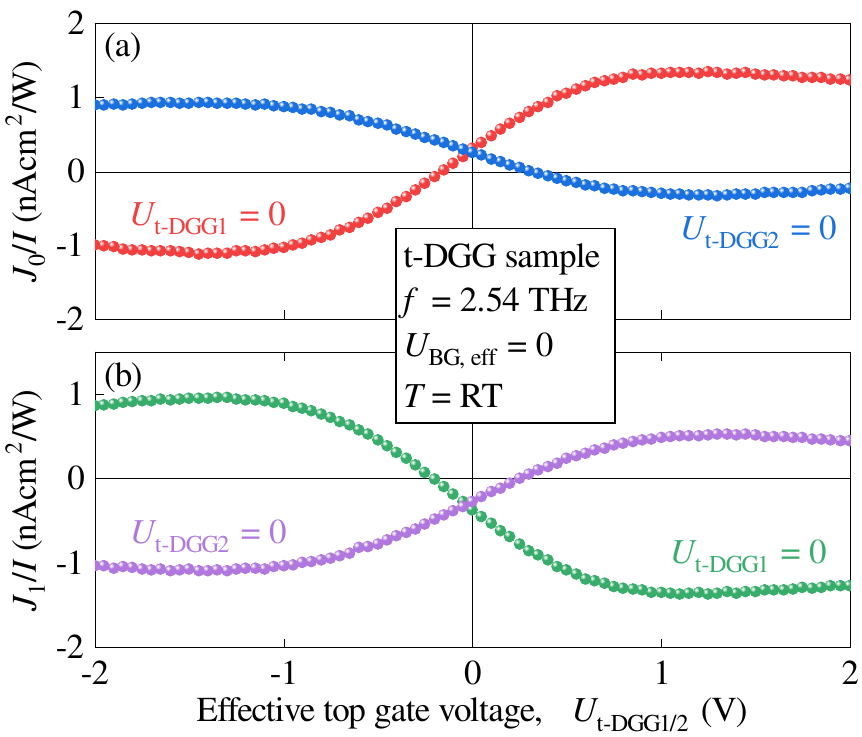}
	\caption{Normalized ratchet photocurrent contributions $J_0$ and $J_1$ as a function of the top gate voltages $U_{\rm t-DGG1/2}$. The data are obtained at room temperature by varying the bias voltage applied to one of the subgate while keeping the other at zero bias. The individual contributions $J_0$ and $J_1$  were extracted from the total photocurrent exploiting the differences of their polarization dependencies, see Eq.~\ref{linear}. Panel (a) shows the polarization independent contribution according to $J_0 = [J (\alpha = 0) + J(\alpha = 90^\circ)]/2$. In Panel (b), the polarization dependent ratchet term is given by $J_1 = [J (\alpha = 0) - J (\alpha = 90^\circ)]/2$. Note that in these measurements the parameter $J_2$ is negligible.}
	\label{Top_gate_dep}
\end{figure}

We start with the data obtained at low-power excitation for a frequency of $f = 2.54$~THz.
Figure \ref{lin_pol_dep}(a) shows the photoresponse normalized to the radiation intensity obtained as a function of the azimuth angle $\alpha$ measured at room temperature labeled in the following as RT. The polarization dependence is exemplarily shown for unbiased back- and subgates of the top DGG. The data points can be fitted well by
\begin{equation}
J = J_0 + J_\mathrm{1}\cos2\alpha + J_\mathrm{2}\sin2\alpha.
\label{linear}
\end{equation}  
This behavior is characteristic for the THz excited ratchet currents investigated in DGG structures, see Refs.~\cite{Faltermeier2015, Ganichev2017, Olbrich2016, Moench2022}. Here, $J_0$ corresponds to the polarization independent part, which is called the Seebeck ratchet effect, whereas the second and third terms are two contributions to the polarization dependent linear ratchet, for review see Ref.~\cite{Ivchenko2011}. A fingerprint of the ratchet effect is the photocurrent dependence on the asymmetry parameter $\Xi$. For zero top gate voltages the photoresponse is due to the non-zero $\Xi$ stemming from the asymmetry of the intrinsic built-in potential, which results from the metal stripes of different widths on top of the encapsulated BLG. In DGG structures the parameter $\Xi$ can also be controllably varied by applying bias voltages, $U_{\rm t-DGG1}$ and $U_{\rm t-DGG2}$, to the individual subgates t-DGG1 and t-DGG2. This is clearly demonstrated in Fig.~\ref{Top_gate_dep}, where one top gate voltage was continuously changed while the other was kept at zero voltage. Panel (a) shows the polarization independent contribution, $J_0$, while panel (b) shows the polarization dependent ratchet response, $J_1$. Here and in the following we focus on the ratchet contributions $J_0$ and $J_1$, since $J_2$ is more than one order of magnitude smaller. Figure~\ref{Top_gate_dep} clearly shows that the variation of an individual subgate voltage, and, consequently, the variation of the lateral asymmetry parameter $\Xi$, changes the sign of the photocurrent sign in the vicinity of zero bias, and increases the photocurrent magnitude with increasing subgate voltage.

In addition to the Seebeck and the linear ratchet currents we detected the circular ratchet effect, which is manifested by the opposite sign of the photocurrent  induced by the right- ($\sigma^+$) and left-handed ($\sigma^-$) circular polarization where the Seebeck ratchet remains constant and the linear one vanishes. Figure~\ref{lin_pol_dep}(b) shows the dependence of the photoresponse on the rotation angle $\varphi$ at room temperature measured at a frequency of $f = 2.54$~THz and keeping all gates at zero bias. The opposite helicities are marked by vertical arrows. The data points can be fitted well by
\begin{equation}
J =  J_0 + \frac{J_1(\cos4\varphi + 1)}{2} + \frac{J_2\sin4\varphi}{2} + J_{\rm circ}\sin2\varphi.
\label{circular}
\end{equation}
Here, $J_{\rm circ}$ determines the amplitude of the circular ratchet effect. It is important to point out that the magnitudes $J_0$, $J_1$, and $J_2$ are the same as in Eq.~\ref{linear}, and the polarization dependent part corresponds to the transformation of the Stokes parameters~\cite{Saleh1991} corresponding to the $\lambda$/4 plate setup~\cite{Belkov2005}.

\section{High-power induced ratchet currents}
\label{highpower}

\begin{figure}
	\centering
	\includegraphics[width=\linewidth]{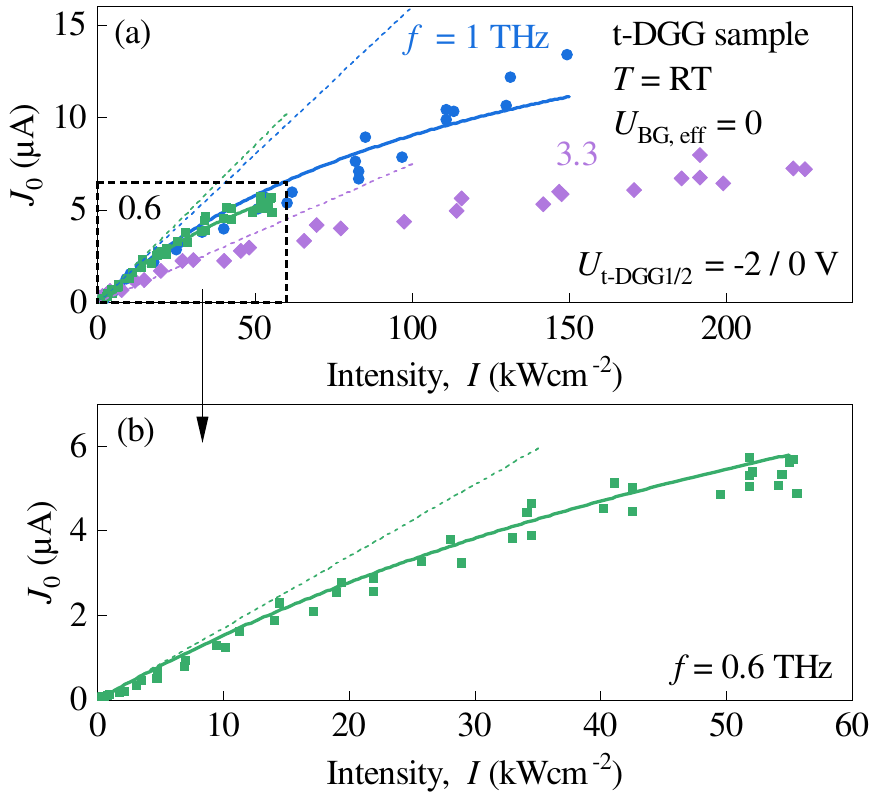}
	\caption{Intensity dependencies of the polarization independent ratchet current $J_0$ measured at room temperature with zero effective back gate voltage $U_{\rm BG} =  0$, $U_{\rm t-DGG1} = -2$~V and  $U_{\rm t-DGG2} = 0$. Panel (b) shows an enlargement of the data at $f$ = 0.6~THz, where the intensity of our laser system is limited to $\approx 60$~kW\,cm$^{-2}$. The solid curves are fits according to Eq.~\ref{saturation_fit} with two fitting parameters: the low-power amplitude $A$ and the saturation intensity $I_s$. These parameters are plotted in Fig.~\ref{Fig7_2} as a function of the excitation frequency. 
	}
	\label{FigRT3}
\end{figure}

\begin{figure}
	\centering
	\includegraphics[width=\linewidth]{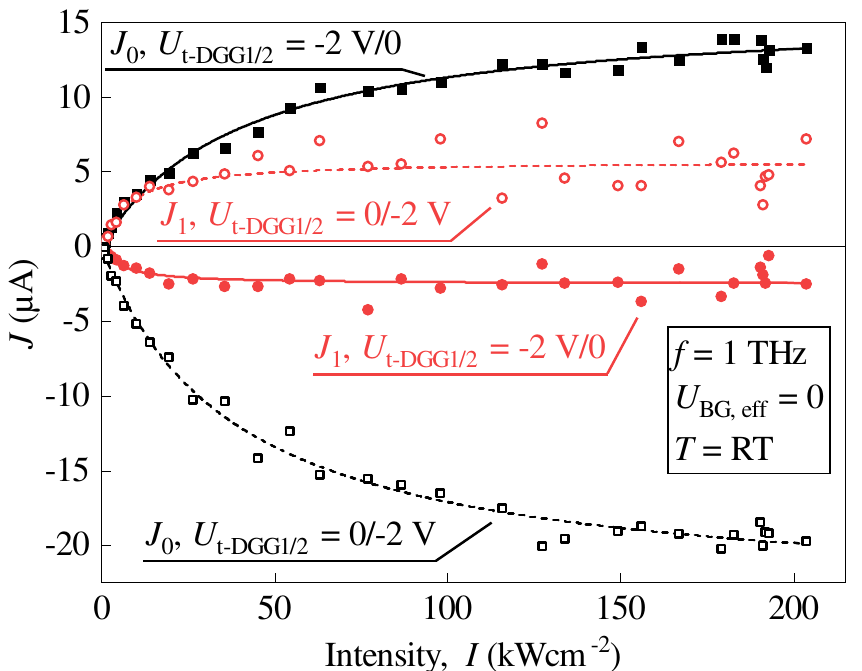}
	\caption{Intensity dependence of the polarization independent ratchet current $J_0$ and the linear ratchet current $J_1$ measured at room temperature and $f = 1$~THz. The data are presented for zero effective back gate voltage $U_{\rm BG} =  0$ and two combinations of top gate voltages. The solid and dashed lines are fits based on Eq.~\ref{saturation_fit}. Note that for the solid and dashed lines the top gate voltages $U_{\rm t-DGG1}$ and $U_{\rm t-DGG2}$ are reversed.
		The fitting parameters are: for $J_0$ at $U_{\rm t-DGG1/2} = -2/0$~V, $A =0.39$~nA/(W/cm$^2$) and $I_s= 40.1$~kW/cm$^2$ (black solid line); for $J_0$ at $U_{\rm t-DGG1/2} = 0/-2$~V, $A =0.62$~nA/(W/cm$^2$) and  $I_s= 38.1$~kW/cm$^2$ (black dashed line); for $J_1$ at $U_{\rm t-DGG1/2} = -2/0$~V, $A =-0.47$~nA/(W/cm$^2$) and  $I_s= 5.3$~kW/cm$^2$ (red solid line); for $J_1$ at $U_{\rm t-DGG1/2} = 0/-2$~V, $A =0.74$~nA/(W/cm$^2$) and  $I_s= 7.6$~kW/cm$^2$ (red dashed line).
		}
	\label{FigRT3_new}
\end{figure}

\begin{figure}
	\centering
	\includegraphics[width=\linewidth]{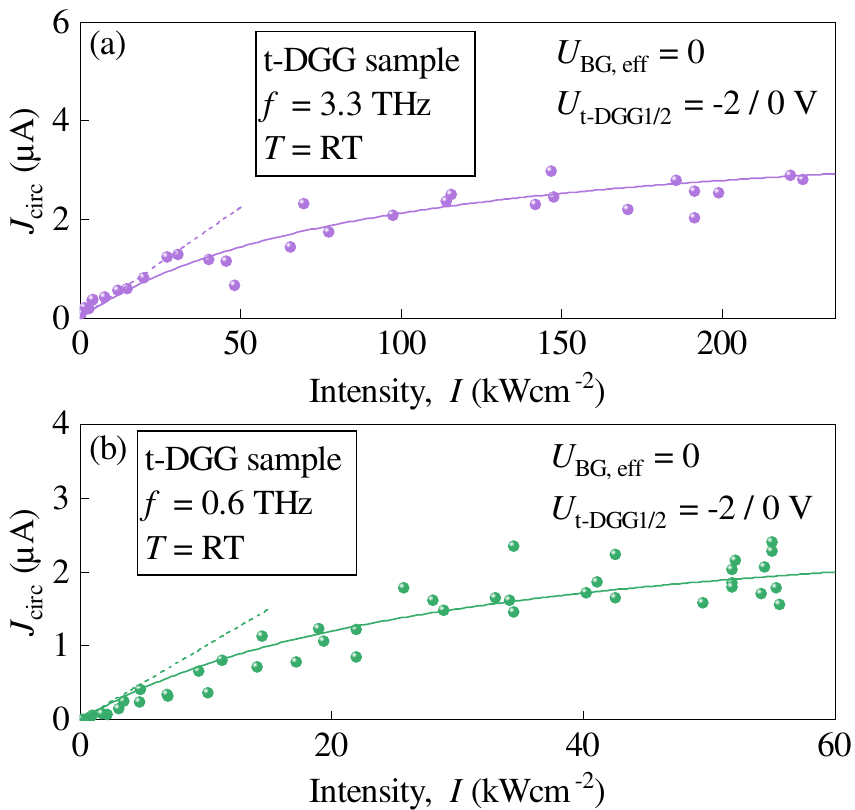}
	\caption{Intensity dependence of the circular ratchet current $J_{\rm circ}$ measured at room temperature with zero effective back gate voltage $U_{\rm BG, eff} =  0$, $U_{\rm t-DGG1} = -2$~V and  $U_{\rm t-DGG2} = 0$. Panels (a) and (b) show the data for the radiation frequency 3.3 and 0.6~THz, respectively.
	The solid curves are calculated according to Eq.~\ref{saturation_fit} with two fitting parameters: the low-power amplitude $A$ and the saturation intensity $I_s$. These parameter are plotted in Fig.~\ref{Fig7_2} as a function of the frequency. The dashed lines represent linear fits of the low-power photocurrent. 
	}
	\label{circ_angle_dep}
\end{figure}

\begin{figure}
	\centering
	\includegraphics[width=\linewidth]{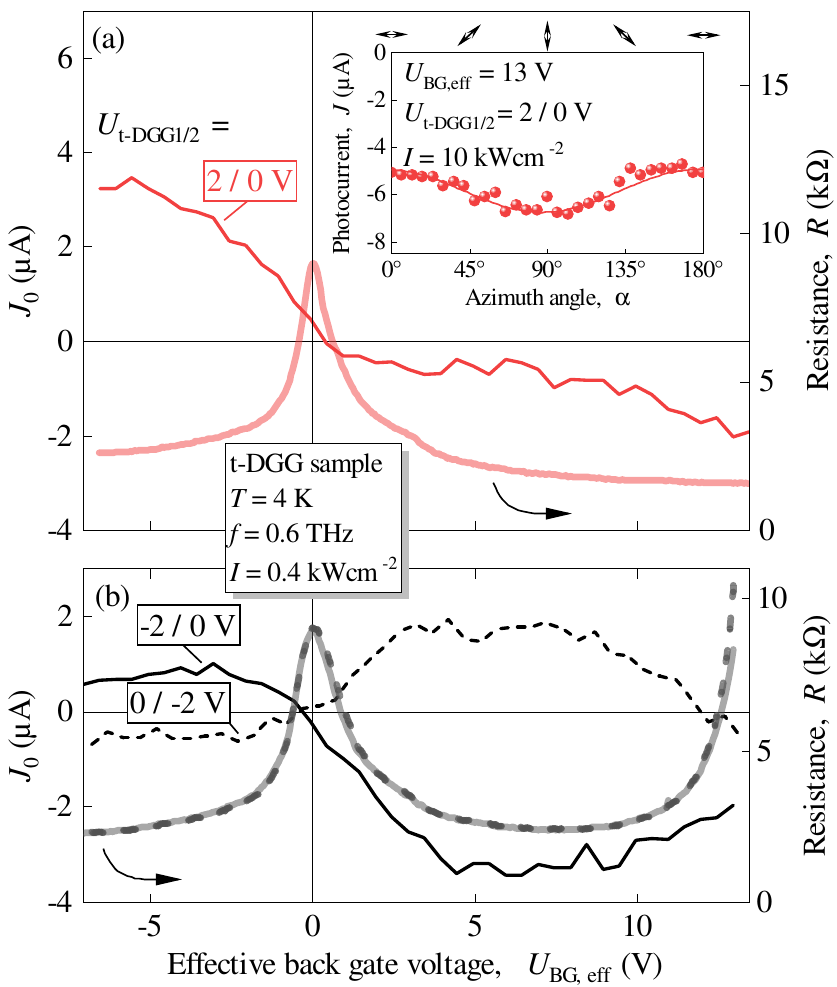}
	\caption{(a, b) Dependence of the photocurrent contribution $J_0$ on the effective back gate voltage $U_{\text{\text{BG, eff}}}$ measured at the radiation frequency of \TeraHertz{0.6}, $T = 4$~K and different top gate voltage combinations $U_{\rm t-DGG1} / U_{\rm t-DGG2}$. The data were obtained at a moderate radiation intensity of 0.4~kW\,cm$^{-2}$.
		The thick lines in the corresponding colors depict the sample resistance. It's worth noting that the resistance curves (solid and dashed) in panel (b) almost coincide. Here, the solid and dashed resistance curves correspond to the $U_{\rm t-DGG1} / U_{\rm t-DGG2} = -2/0$~V and 0/-2~V, respectively. The inset shows the dependence of the photocurrent on the azimuth angle $\alpha$ obtained for a radiation intensity of 10~kW\,cm$^{-2}$ for gate voltage combinations of $U_{\text{BG,eff}}=\SI{11.2}{\volt}, $ $U_{\text{t-DGG1}}=\SI 2{\volt}$, and $U_{\text{t-DGG2}}=\SI{0}{\volt}$. The red solid curve is a fit according to Eq.~\ref{linear} with fitting parameters, $J_0 = -5.85$~$\mu$A, $J_1 = 0.9$~$\mu$A and $J_2 = -0.1$~$\mu$A. Double arrows on the top illustrate the state of polarization for several values of the azimuth angles $\alpha$.
		}
	\label{Fig4_2}
\end{figure}

The results described so far have been obtained by applying low-power radiation ranging in fractions of W/cm$^2$. At these low intensities the photocurrent linearly increases with the radiation intensity (not shown). By using the pulsed laser system, we have been able to increase the radiation intensity by up to five orders of magnitude, i.e.,  up to hundreds of kW/cm$^2$. Such a drastic increase in the radiation intensity results in a nonlinear behavior of the photocurrent. Measuring the photocurrent at room temperature we found that it saturates with the increase of the radiation intensity $I$. Figure~\ref{FigRT3} shows the intensity dependence of the ratchet current $J_0$ obtained for negatively biased t-DGG1 and zero biased t-DGG2. This sequence of subgate voltages yields a high asymmetry parameter $\Xi$ and hence a high ratchet current, as shown in Fig.~\ref{Top_gate_dep}. The data can be fitted well by an empirical equation 
\begin{equation}
	\label{saturation_fit}
	J=A\cdot\frac{I}{1+I/I_s}\,,
\end{equation}
where the amplitude $A$ and the saturation intensity $I_s$ are used as fitting parameters. The parameter $A$ describes the photocurrent amplitude in the linear regime ($I \ll I_s$). The magnitude of $A$, detected in pulsed measurements at frequencies 1 and 3.3~THz, is close to that obtained in low-power $cw$ measurements at a frequency of 2.54~THz, see Figs.~\ref{FigRT3}(a) and \ref{Top_gate_dep}(a), respectively. The data show that the $A$ decreases with increasing frequency, while the saturation intensity $I_s$ increases with increasing $f$. This behavior is discussed in detail in Sec.~\ref{discussion}. Saturation is also observed for the linear ratchet effect $J_1$, see Fig.~\ref{FigRT3_new}. This figure also shows that the $J_0$ and $J_1$ contributions always have opposite signs. As expected for the ratchet current, the sign of these contributions reverses when the gate combination is inverted. Applying $\sigma^+$ and $\sigma^-$ polarized radiation we observed that the intensity dependence of the circular ratchet current is also well described by Eq.~\ref{saturation_fit}. The corresponding data and the fits are shown in Fig.~\ref{circ_angle_dep}. The circular photocurrent was calculated as half the difference between the photocurrents excited by right- and left-handed circularly polarized radiation.

\begin{figure}
	\centering
	\includegraphics[width=\linewidth]{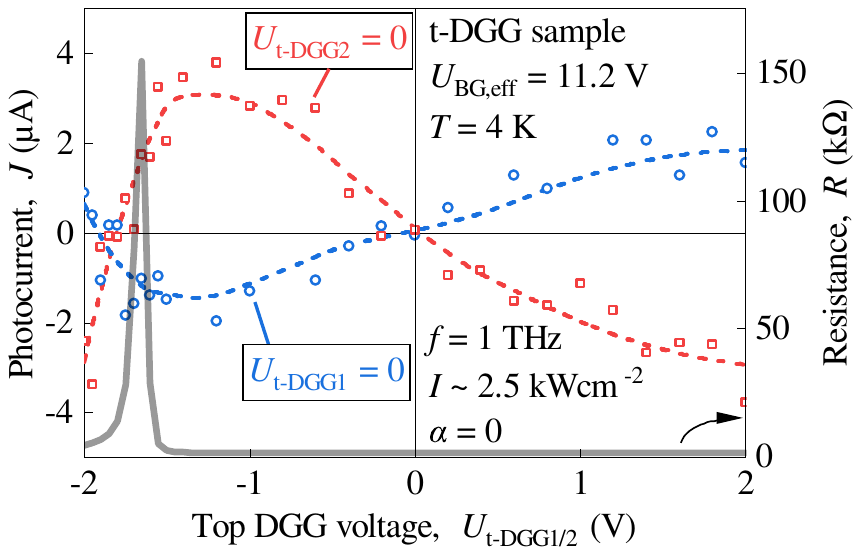}
	\caption{
		Dependencies of the photocurrent on the top gate voltages $U_{\text{\rm t-DGG1}}$ (red squares) and  $U_{\text{\rm t-DGG2}}$ (blue circles) for a radiation intensity of 2.5~kW\,cm$^{-2}$, radiation frequency of \TeraHertz{1}, and a back gate voltage of $U_{\text{BG,eff}}=\SI{11.2}{\volt}$. To measure the dependence on $U_{\text{\rm t-DGG1}}$ ($U_{\text{\rm t-DGG2}}$) the other top gate voltage was set to zero as indicated near the curves. The red and blue dashed lines are introduced as a guide for the eye. The thick gray line depicts the corresponding sample resistance as a function of $U_{\text{\rm t-DGG1}}$, while the $U_{\text{\rm t-DGG2}}$ was kept unbiased.
	}
	\label{Fig2}
\end{figure}

\begin{figure}
	\centering
	\includegraphics[width=\linewidth]{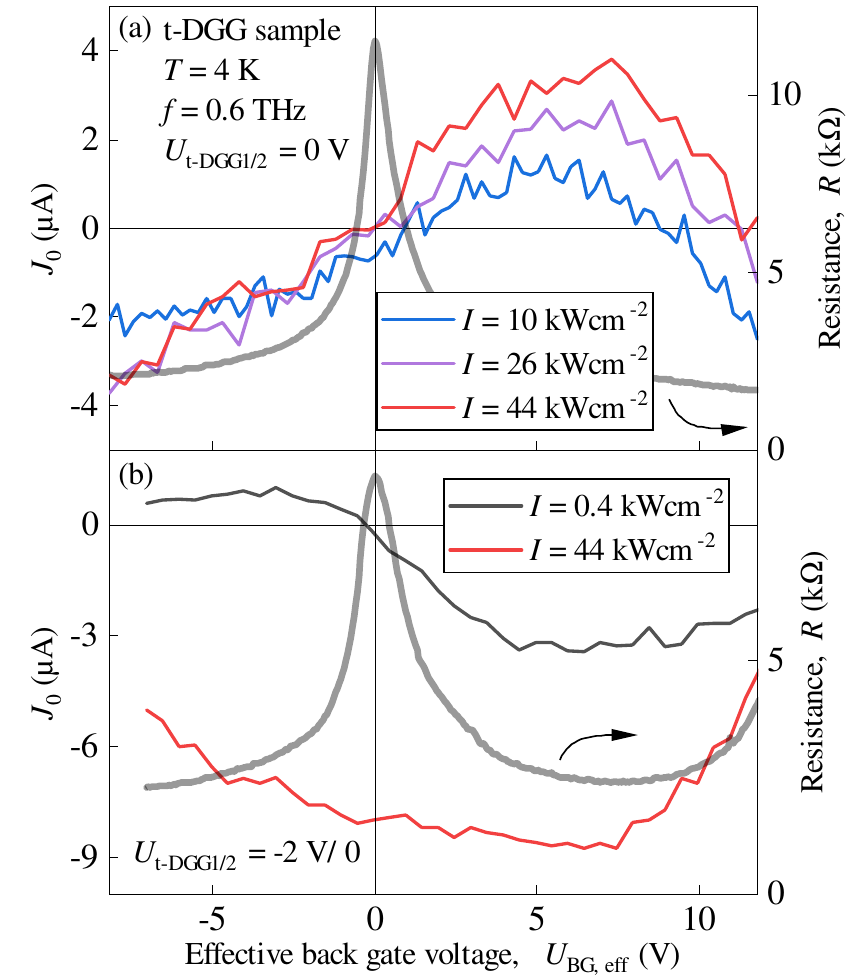}
	\caption{Dependencies of the photocurrent contribution $J_0$ on the effective back gate voltage $U_{\text{\text{BG, eff}}}$ measured for a radiation frequency of \TeraHertz{0.6} and different radiation intensities. Panel (a) shows the data obtained for zero bias at both top gates. , while panel (b) displays the data for $U_{\text{t-DGG1}}= -2$~V, keeping t-DGG2 unbiased. The thick gray lines on both panels represent the corresponding sample resistance.
	}
	\label{Fig4_1}
\end{figure}

\begin{figure}
	\centering
	\includegraphics[width=\linewidth]{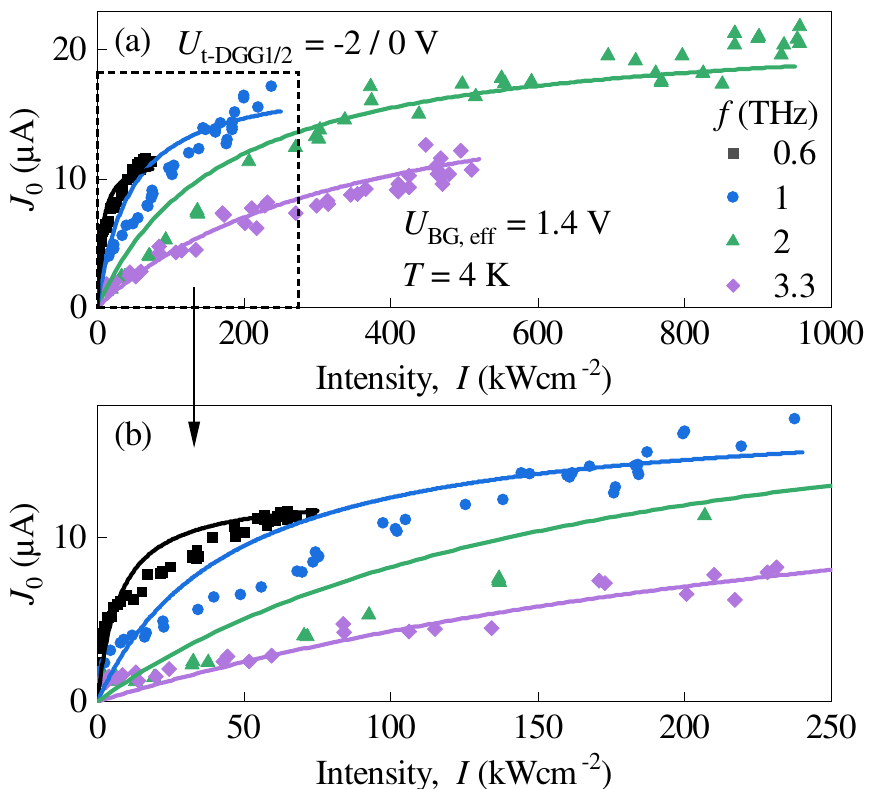}
	\caption{Dependencies of the photocurrent contribution $J_0$
	for the top gate voltage combination $U_{\text{TGG1/2}}= \SI{-2}{\volt} / 0$. The data are obtained for several radiation frequencies and an effective back voltage of \SI{11.2}{\volt}. Panel (b) presents an enlargement of the data for $f = 0.6$~THz and $f = 1$~THz, where the intensity of our laser system is limited to $\approx 60$~kW/cm$^2$ and 250~kW\,cm$^{-2}$, respectively. Solid curves are fits based on Eq.~\ref{saturation_fit} with the low-power amplitude $A$ shown in Fig.~\ref{Fig7_2}. Note that the fits reflect the trend but describe the data for moderate intensities in a rough way. The saturation parameter $I_s$ used for these fits is about 10~kW\,cm$^{-2}$ for $f = 0.6$~THz and increases with frequency as $I_s \propto \omega^2$.
}
	\label{Fig5}
\end{figure}

\begin{figure}
	\centering
	\includegraphics[width=\linewidth]{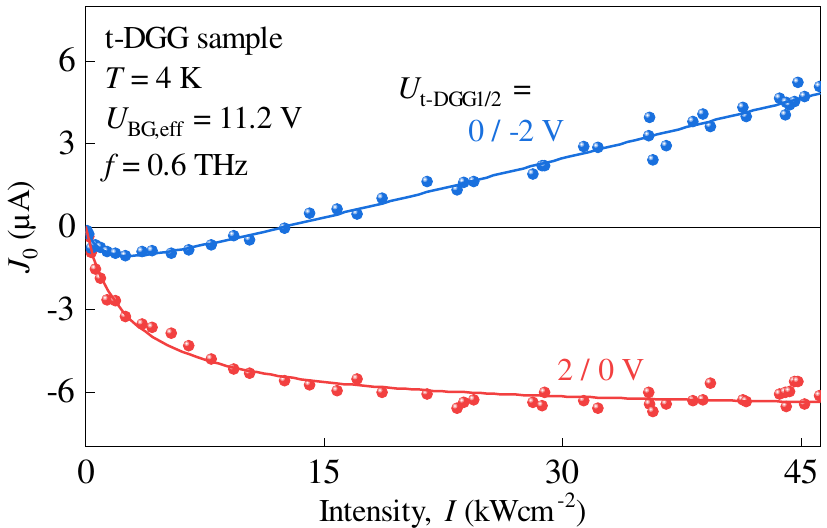}
	\caption{
		Intensity dependencies of the photocurrent $J_0$ obtained for a radiation frequency of \TeraHertz{0.6}, an effective back gate voltage of \SI{11.2}{\volt} and different combinations of the top subgate bias voltages.  The solid curves are fits according to Eq.~\ref{signchange_fit}. 
	The fitting parameters for: 
	the blue curve are A = \fittingunit{-5.95} $I_{s,A}=\kWcm{0.9}$, $B = \fittingunit{0.11}$, $I_{s,B} = \kWcm{3E22}$
	and the red curve are
	A = \fittingunit{-1.9} $I_{s,A}=\kWcm{0.5}$, $B = \fittingunit{-1.6}$, $I_{s,B} = \kWcm{3.6}$.
}
			\label{Fig5_2}
\end{figure}

Now we focus on the data obtained at $T = 4$~K. 
	 The inset in Fig.~\ref{Fig4_2} shows the polarization dependence of the photocurrent excited by linearly polarized radiation of \textit{moderate} intensity. 
	 In spite of that fact that this dependence is similar to that obtained at room temperature and is well described by Eq.~\ref{linear}, at $T = 4$~K the photocurrent is mainly dominated by the polarization independent contribution $J_0 \gg J_1, J_2$. Therefore, below we concentrate on examining the functional behavior of $J_0$. 
	 
Figure~\ref{Fig4_2} shows the dependencies of $J_0$ on the effective back gate voltage for different subgate voltages combinations. All the curves change their sign near the CNP (see corresponding sheer transport curves).
The sign change of the photocurrent results from the sign change of the majority carriers at the CNP. The functional dependencies of the ratchet currents are odd with respect to the carrier charge, so a transfer from hole to electron conductivity leads to an inversion of the ratchet current direction. 
%
The data presented in Fig.~\ref{Fig4_2} demonstrate, that
the curves with inverted subgate voltages have consistently opposite signs, see solid and dashed curves for $U_{\rm t-DGG1} / U_{\rm t-DGG2} = -2~{\rm V}/ 0$ and $U_{\rm t-DGG1} / U_{\rm t-DGG2} = 0 / -2~{\rm V}$ in Fig.~\ref{Fig4_2}(b). Opposite ratchet current directions for these subgate voltage combinations originate in the sign change of the lateral potential gradient $dV/dx$ and, consequently, the asymmetry parameter $\Xi$ (see Eq.~\ref{Xi0}).
%
%
Figure~\ref{Fig4_2}(b) illustrates that when a high negative bias voltage is applied to the subgates ($U_{\rm t-DGG1/2}$), the ratchet current behaves in a non-monotonic way: it increases with $U_{\rm BG, eff}$, reaches an extreme, and then decreases or even changes the sign (as shown by the solid and dashed black curves). Studying the resistance of the samples we found that, apart from a maximum at the CNP, it increases drastically at high negative $U_{\rm t-DGG1/2}$, see black lines in Fig.~\ref{Fig4_2}, i.e. for the gate voltages at which the peculiarity of the ratchet current is detected. This additional increase can be understood qualitatively in an extended capacitor model, see Refs.~\cite{Castro2010,Oostinga2007}. While the back gate acts on the entire graphene flake, the top gates only couple to the graphene regions underneath, shifting the CNP only in those regions. If the top gates are biased, the carrier concentration in the regions directly below is changed, resulting in an additional resistance peak in the back gate sweep. 

A double sign inversion of the ratchet current is also found by studying the top gate voltage dependencies, see  Fig.~\ref{Fig2}.
Here, the photocurrent exhibits two characteristic sign changes: 
One of them takes place around zero top gate voltage, while the second sign change is observed around the CNP. While the sign inversion  of the photocurrent in the vicinity of the CNP is discussed above, and corresponds to the change from positively charged holes to negatively charged electrons, 
the sign inversion around zero top gate voltage is caused by the reversal of  the lateral potential gradient $dV/dx$. A consistently opposite sign of the traces obtained by varying $U_{\rm t-DGG1}$ and $U_{\rm t-DGG2}$, see red and blue traces in Fig.~\ref{Fig2}, is also caused by the sign variation of  $\Xi$. Indeed, biasing the wide (t-DDG1) while keeping the narrow (t-DGG2) fingers at zero bias results in an opposite sign of $\Xi$ with respect to the configuration in which t-DGG1 is is held at zero potential and t-DGG2 is biased. In fact, in these configurations we have an opposite sign of $dV/dx \propto \Xi$.
		%
Note that this peculiar dependence on the asymmetry parameter is a fingerprint of the ratchet effect.

Increasing the radiation intensity substantially changes the effective back gate dependencies. Figure~\ref{Fig4_1} exemplarily shows $J_0$ as a function of $U_{\rm BG, eff}$ obtained for different radiation intensities and two sequences of top subgate voltages. The figure shows a highly nonlinear behavior of the ratchet current, which varies with back and top gate voltages. For example Fig.~\ref{Fig4_1}(a) shows that at negative back gate voltages and zero biased top gates the signal remains almost unchanged by the increasing of $I$ by changing the intensity from 10 to 44~kW/cm$^{2}$, while for high positive back gates an increase in intensity even changes the sign of $J_0$. For  $U_{\rm t-DGG1} / U_{\rm t-DGG2} = -2~{\rm V}/ 0$ the situation reverses, see Fig.~\ref{Fig4_1}(b), now the sign change with increasing intensity is clearly seen at negative gate voltages and the ratchet current saturates at high positive gate voltages. To study these nonlinearities we measured the intensity dependencies for different sequences of gate voltages.

We begin with the gate voltages at which the current saturates with increasing $I$. Figure~\ref{Fig5} shows the photocurrent measured at different frequencies, low back gate voltage $(U_{\rm BG, eff}=1.4$~V) and top subgate voltages equal to $U_{\rm t-DGG1} / U_{\rm t-DGG2} = -2~{\rm V}/ 0$. The ratchet current follows the fit curves calculated using the empirical formula, see Eq.~\ref{saturation_fit}, used above for the room temperature data. These fits describe roughly the intensity dependencies at moderate intensities. 
Compared to the room temperature data, see Fig.~\ref{FigRT3}, the amplitude of the unsaturated current ($I << I_s$) at $f = 0.6$~THz increases by an order of magnitude and shows an even stronger reduction with increasing frequency. This behavior will be discussed in detail in Sec.~\ref{discussion}. Strikingly, for a particular set of top gate voltages we found that Eq.~\ref{saturation_fit} does not apply at all. Several examples are shown in Fig.~\ref{Fig5_2} depicting $J_0(I)$ measured at a high back gate voltage $(U_{\rm BG, eff}=11.2$~V) and three subgate combinations. It demonstrates the nonmonotonic dependencies of the ratchet current and even changes sign with increasing radiation intensity, see traces for $U_{\rm t-DGG1} / U_{\rm t-DGG2} = -2~{\rm V}/ 0$ and $U_{\rm t-DGG1} / U_{\rm t-DGG2} = 0/-2~{\rm V}$. We found that the data can be well fitted by an empirical formula
\begin{equation}
\label{signchange_fit}
J=A\cdot\frac{I}{1+I/I_{s,A}}+B\cdot\frac{I}{1+I/I_{s,B}}\,,
\end{equation}
with amplitudes $A$ and $B$ and the corresponding saturation intensities $I_{s,A}$ and $I_{s,B}$ as fitting parameters, see Fig.~\ref{Fig5_2}. This indicates that different mechanisms are involved in the generation of the photocurrent.  Below we show that a two mechanism assumption valid for the whole set of low-temperature data and indeed is in line with the microscopic theory. However, the dependence of the functional behavior is more complex and one must consider that the parameter $A$ is $I$-dependent.

\begin{figure}
	\centering
	\includegraphics[width=\linewidth]{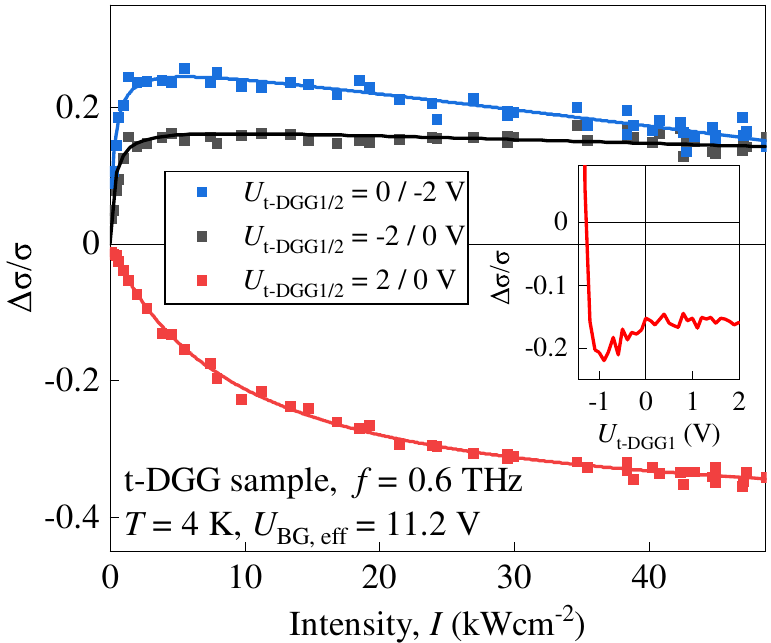}
	\caption{
		Dependence of the photoconductivity signal on the radiation intensity for different gate voltage combinations and an effective back gate voltage of \SI{11.2}{\volt}. 
	The solid lines are a guide for the eye. 
The inset shows the dependence of the $\Delta\sigma/\sigma$ on the top gate voltage  $U_{\text{\rm t-DGG1}}$. 	
	}
	\label{Fig8}
\end{figure}

\section{THz photoconductivity}
\label{photoconductivity}

Further analysis of the intensity dependence, see Sec.~\ref{discussion}, shows that the radiation induced electron gas heating and, a consequent change of the conductivity of the sample should be taken into account to describe both the saturation processes and the intensity dependent amplitude $A$.

 Therefore, we performed additional measurements of the photoinduced change of the sample's conductivity $\Delta \sigma$ (bolometric photoconductivity). Figure~\ref{Fig8} shows the intensity dependence of the normalized photoconductivity $\Delta\sigma/\sigma$ obtained under the same conditions as the photocurrent data presented in Fig.~\ref{Fig5_2}. The data show that depending on the combination of the top subgate voltages the photoconductivity is either positive (conductivity increases upon heating, $\Delta\sigma/\sigma >0$), see traces for $U_{\rm t-DGG1} / U_{\rm t-DGG2} = 0 / -2~{\rm V}$ and  $U_{\rm t-DGG1} / U_{\rm t-DGG2} =   -2~{\rm V}/ 0$, or negative (conductivity decreases upon heating,  $\Delta\sigma/\sigma <0$). The sign inversion is also clearly present in the dependence of  $\Delta\sigma/\sigma (U_{\rm t-DGG1})$, see the inset in Fig.~\ref{Fig8}. All these observations clearly show a substantial radiation-induced electron gas heating which, under certain conditions, results even in a change of the scattering mechanisms, see inset in Fig.~\ref{Fig8}. Furthermore, Fig.~\ref{Fig8} demonstrates that the photoconductivity depends nonlinearly on the radiation intensity. It saturates at high positive and negative $U_{\rm t-DGG1}$, while for a high negative $U_{\rm t-DGG2}$ the behavior becomes nonmonotonic approaching its maximum at about 5~kW\,cm$^{-2}$. Below we will discuss this behavior in more detail.

\section{Discussion}
\label{discussion}

\subsection{Low-power results}

The theoretical analysis performed in Refs.~\cite{Olbrich2016,Moench2022} shows that the photocurrent is caused by a combined action of a spatially periodic in-plane potential and the radiation spatially modulated due to near-field effects of the diffraction on the DGG stripes. The ratchet effect is controlled by the lateral asymmetry parameter $\Xi \propto dV(x)/dx$, see Eq.~\ref{Xi0}. In the top DGG devices this parameter can be controllably varied by applying the voltages $U_{\rm t-DGG1}$ and $U_{\rm t-DGG2}$ to the individual subgates. Note that, switching the gate voltage from $U_{\rm t-DGG1} > 0$, $U_{\rm t-DGG2} = 0$ to $U_{\rm t-DGG1} = 0$, $U_{\rm t-DGG2} > 0$ leads to a change in sign of $\Xi$  and, as a consequence, to a reversal of the photocurrent direction, as illustrated in Fig.~\ref{Top_gate_dep}.  This confirms that the photocurrent is caused by the ratchet effect. 

\begin{figure}
	\centering
	\includegraphics[width=\linewidth]{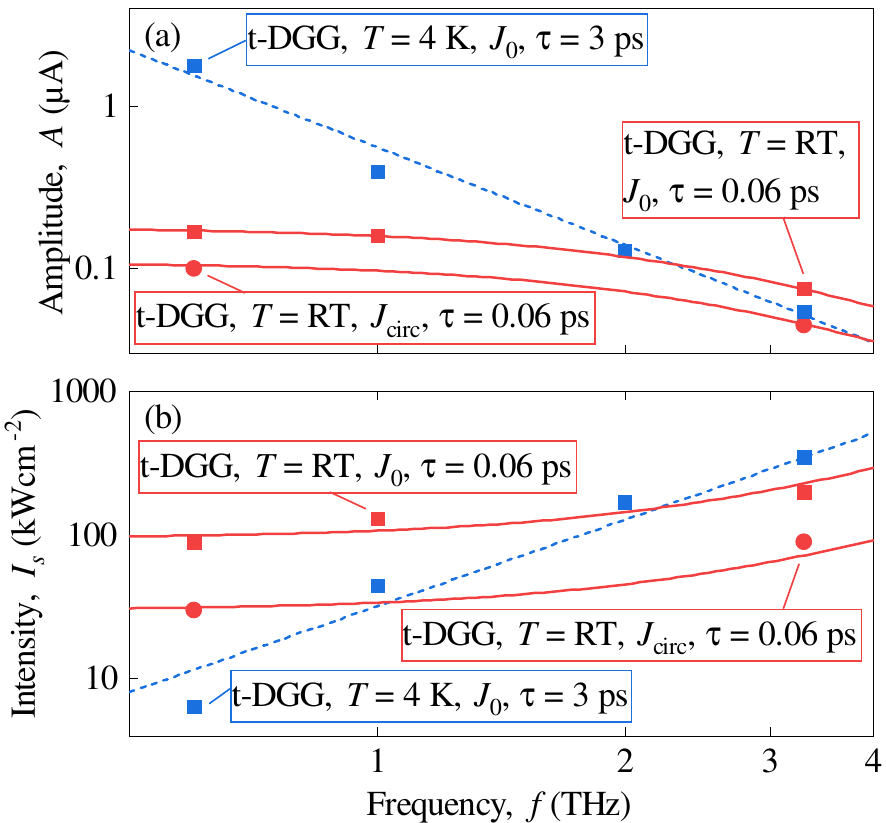}
	\caption{
		Frequency dependencies of the low-power ratchet current amplitude $A$, panel (a), and the saturation intensities $I_s$, panel (b), measured in the top DGG device. Solid and dashed lines show fits according to $J \propto A \propto 1/(1+\omega^2\tau^2)$ (a) and $I_s  \propto (1+\omega^2\tau^2)$ (b). The  momentum relaxation times $\tau$ obtained from transport measurements, and the sample's temperature are indicated close to each curve. 
	}
	\label{Fig7_2}
\end{figure}

It is known that the ratchet effect can have three contributions: the polarization-independent ratchet current $J_0$, the one defined by the relative orientation of the electric field vector $\bm E$ and the DGG stripes (linear ratchet, $J_1$), and the circular ratchet, $J_{\rm circ}$ the direction of which is defined by the radiation helicity and inverses by changing from right- to left- circularly polarized radiation. All three contributions have been detected in both top and bottom DGG devices. The corresponding polarization dependencies are shown in Fig.~\ref{lin_pol_dep},
 \ref{FigRT1_alpha}, and \ref{FigRT1_circ}.

In the THz range the ratchet currents are typically caused by the indirect intraband optical transitions (Drude-like absorption)~\footnote{At low temperature and close to the CNP, direct interband optical transitions may also contribute to the ratchet current}
which scale with the radiation frequency as $\eta^{\rm Drude} \propto 1/(1+\omega^2\tau^2)$. The decrease of the ratchet current amplitude with increasing frequency is observed in both types of devices at room temperature and $T = 4$~K. Our results reveal that the current amplitude in the bottom DGG device is about 10 times smaller than that in the top DGG, therefore, in the following we focus on the latter device. The frequency dependence of the low-power ratchet current amplitude is shown in Fig.~\ref{Fig7_2} together with the fits after $J \propto  A \propto  1/(1+\omega^2\tau^2)$. The momentum relaxation times used for the fits correspond to the transport momentum relaxation times measured in the structures studied in Ref.~\cite{Moench2023}. The fits demonstrate that the ratchet current is well described by the frequency dependence of the Drude absorption.

The above conclusions, which are general for the ratchet photocurrents, are also valid for the ratchet currents obtained in the bottom DGG device, as discussed in Appendix~\ref{bottomDGG}. However, the lateral asymmetry parameter $\Xi$ and, consequently the ratchet current in the novel device are additionally controlled by the back gate voltage.
This is because the applied back gate voltage is periodically screened by the conductive DGG graphite stripes, which 
results in a periodic lateral potential and therefore affects
the parameter $\Xi$. 

\subsection{High-power results}
\label{high_power_top_gate}

Increasing the power of the radiation by five orders of magnitude did not change the functional behavior of the THz-induced ratchet current. In fact, at high power the characteristic ratchet current behavior, such as $J \propto \Xi$, back gate and the polarization dependencies, remain qualitatively unchanged. This is shown for the top DGG structure in Figs.~\ref{FigRT3_new} and \ref{Fig4_2} - \ref{Fig4_1}.

%
 An important difference, however, is that the ratchet current magnitude behaves nonlinearly as the radiation intensity increases, see Figs.~\ref{FigRT3} - \ref{circ_angle_dep}, and \ref{Fig4_1} - \ref{Fig5_2}.
 In all cases the signal initially increases linearly with the radiation intensity, and subsequently either saturates (in most cases) or approaches maximum and changes sign. We note that the magnitude of the linear-in-$I$ amplitude of the ratchet current corresponds to that of the $cw$ low-power THz laser and exhibits the frequency dependence which is characteristic for the Drude absorption, see Fig.~\ref{Fig7_2}(a). Therefore, we will focus on the intensity dependence of the photocurrent. 

First we discuss the results obtained in 
devices at room temperature. Figures~\ref{FigRT3} and \ref{circ_angle_dep} 
show that at room temperature  the ratchet current grows with increasing intensity $I$ and saturates at high intensity. The overall intensity dependence can be well fitted by Eq.~\ref{saturation_fit} with saturation intensity $I_s$ depending on the radiation frequency $f$. 

Analysis of the fit functions showed that the saturation intensity changes as $I_s  \propto (1+\omega^2\tau^2)$ and that the values of the relaxation time $\tau$ are those used for the fits of the ratchet current amplitude $A \propto \eta^{\rm Drude}$. This is shown for the top DGG device in Fig.~\ref{Fig7_2}(b). 
In the frequency range studied, $I_s$ increases from 10 to 500 kW\,cm$^{-2}$.
These results indicate that DGG-based detectors of THz radiation have a high dynamic range, remaining linear at room temperature for at least up to 100 kW\,cm$^{-2}$.

The observed saturation of the ratchet current is attributed to the absorption bleaching caused by the electron gas heating. The bleaching of the Drude-like radiation absorption in graphene has recently been observed in experiments on nonlinear ultrafast THz spectroscopy in monolayer graphene~\cite{Mics2015}, as well as for photogalvanic currents in bilayer~\cite{Candussio2021a} and twisted~\cite{Hubmann2022} graphene structures. The range of frequencies (0.4 – 1.2 THz) and radiation electric fields  (2 – 100 kV\,cm$^{-1}$) used in our study are similar to those used in Refs.~\cite{Mics2015, Candussio2021a, Hubmann2022}. These works show that the absorption bleaching is well described by the empirical formula 
	\begin{equation}	
		\eta^{\rm Drude} \propto \left(1+ \frac I{I_{s}}\right)^{-1}\,, \label{eq_absorbance1}
	\end{equation}
where the saturation intensity $I_{s}$ is proportional to the Drude absorption cross-section and the reciprocal energy relaxation time. This proportionality describes the observed frequency dependence of the saturation intensities in both types of devices well, see Fig.~\ref{Fig7_2}(b) and discussion above.

Now we turn to the ratchet photocurrent nonlinearity detected at low temperatures. Intensity dependencies obtained for both types of devices at different combinations of the DGG subgates and back gate voltages are shown in  Figs.~\ref{Fig5} and \ref{Fig5_2}.
The data demonstrate that under most experimental conditions the ratchet current saturates with increasing radiation power. However, the functional behavior of the photocurrent becomes more complex. Figure~\ref{Fig5} shows the intensity dependencies obtained in top DGG devices at $T= 4$~K together with fits based on Eq.~\ref{saturation_fit}. The fits allow us to extract the low intensity current magnitude $A$ and reflect the trend of the signal to saturate. For high intensities the fits agree only roughly with the data. Both the low-power current amplitude $A$ and the occurrence of saturation exhibit a stronger frequency dependence as compared to the room temperature data. Corresponding dependencies for the top DGG structure are shown in Fig.~\ref{Fig7_2}, which are well described by $A \propto \eta^{\rm Drude}\propto 1/(1+\omega^2\tau^2)$ and $I_s \propto (1+\omega^2\tau^2)$ with the momentum relaxation times $\tau = 3$~ps obtained from transport measurements, see Ref.~\cite{Moench2023}.

The saturation of the ratchet photocurrent can also be attributed to electron gas heating. 
For complementary THz photoconductivity results, see Fig.~\ref{Fig8}.
The observed photoconductivity is caused by radiation-induced electron gas heating and the associated mobility reduction. Figure~\ref{Fig8} 
shows that the photoconductivity depends highly nonlinear on the radiation intensity and, is either positive or negative depending on the combination of subgate voltages. Opposite signs of the photoconductivity demonstrate that the electron gas heating decreases the sample's resistivity (positive photoconductivity) or increases its resistivity (negative photoconductivity) revealing the presence of two different scattering mechanisms. This is confirmed by measuring the sample's resistance as a function of the temperature demonstrating that at low temperatures and a certain combination of gate voltages the $dR/dT$ changes its sign with increasing temperature. 
In photoconductance measurements this results in a transition from positive to the negative photoconductivity~\cite{Ganichev2005}. 

This sign change, originating in electron gas heating, solely describes the complex nonlinearity of the polarization independent ratchet current, $J_0$. Indeed, one of the well known mechanism is the Seebeck ratchet current, which stems from an inhomogeneous electron gas heating by the near field $\bm E(x)$ caused by the diffraction of the radiation incident on the lateral superlattice. The electron-phonon scattering followed by the radiation absorption causes the position-dependent electron temperature characterized by the periodic profile $\delta T(x) \propto |{\bm E(x)}|^2$. According to Refs.~\cite{Nalitov2012,Olbrich2016,Faltermeier2017}, the Seebeck ratchet current density is given by
\begin{equation}
\label{eq:j:Seebeck}
j_x =
-\Xi {{\partial{\sigma}}\over{\partial{T}}}{e\tau\tau_\varepsilon \over m [1+ (\omega\tau)^2]}.
\end{equation}
Here $\Xi$ is, as before, the asymmetry parameter (see Eq.~\ref{Xi0}), $\tau_\varepsilon$ is the electron energy relaxation time and $\sigma(T)$ is the temperature-dependent conductivity. Indeed, on the one hand, the photocurrent is proportional to the Drude absorption, which saturates as the radiation intensity, whereas on the other hand, the photocurrent is proportional to $\partial{\sigma}/\partial{T}$. Consequently, a sign change of the photocurrent shown in Fig.~\ref{Fig5_2} can result from the sign inversion of $\Delta\sigma/\sigma$ with increasing $I$. However, this contradicts the experiment's findings, which demonstrate that the photoconductivity saturates without changing the sign (Fig.~\ref{Fig8}) for appropriate gate voltage combinations where sign inversion is detected (Fig.~\ref{Fig5_2}). 

All these observations show that the sign inversion is caused by the interplay of two different mechanisms of the ratchet current. In fact, another mechanism responsible for current formation has been introduced in Refs.~\cite{Ivchenko2011,Nalitov2012,Faltermeier2018} and is called the dynamic carrier density redistribution (DCDR). Here, an \textit{ac} electric field $\bm E(x)\exp(-i\omega t)+c.c.$ with near-field amplitude $|{\bm E(x)}|$, which is applied to an electronic system with static potential $V(x)$ results in a space-periodic, time-oscillating carrier density profile $\delta N(x,t)=\delta N_\omega(x)\exp(-i\omega t)+c.c.$, see Appendix \ref{DCDR}. Accordingly, the corresponding photocurrent is described by~\cite{Faltermeier2018}
	\begin{equation}
	\label{eq:j_non-heat}
	j_x^{\rm DCDR} = \Xi {e^3\tau^3\over \pi \hbar^2 m [1+ (\omega\tau)^2] } (1+P_L),
	\end{equation}
%
containing both polarization-dependent ($\propto P_L$) and independent parts with equal amplitudes. The equation shows that, like the Seebeck ratchet effect it is proportional to the Drude absorption but in contrast independent of $\partial{\sigma}/\partial{T}$. Consequently, the Seebeck and the DCDR ratchet effects are characterized by different nonlinear contributions. Furthermore, the direction of the corresponding currents are opposite, therefore, an increase of the radiation intensity can change the sign of the current indicating that one mechanism becomes dominant. The interplay of different ratchet current mechanisms together with the nonlinearity of the BLG conductance makes it difficult to describe the ratchet current nonlinearity observed in low temperature measurements analytically.
Nevertheless our data show that the ratchet current behaves linearly with increasing radiation intensity up to couple of kW\,cm$^{-2}$ and more. This fact, together with the substantial increase of the ratchet current magnitude, demonstrates that also He-cooled DGG devices can efficiently be used for the THz radiation detection.

So far, we presented and discussed the results for the top DGG structures. However, our research on the ratchet effect in the bottom DGG (see Appendix~\ref{bottomDGG}) demonstrates that the origin and the nonlinearity of the ratchet current are general, yielding qualitatively similar findings across different device designs.


\section{Summary} 
\label{summary}

In summary, we investigated asymmetric DGG THz ratchet devices based on BLG with two different designs. The ratchet current consists of three contributions: a polarization-independent, a linear and a circular ratchet effect. 

Our findings show that the photoresponse in these structures is linear up to high radiation intensities, ranging from several hundreds of kW\,cm$^{-2}$ at room temperatures and  tens to hundreds kW\,cm$^{-2}$ at 4~K. At higher intensities the nonlinearity occurs due to radiation-induced electron gas heating. For high back gate voltages the current magnitude is proportional to the Drude absorption and, therefore, increases with decreasing frequency as $1/(1+\omega^2\tau^2)$. Under most conditions the ratchet current saturates with increasing radiation intensity. The saturation intensity depends on the bias applied to the DGG subgates as well as on the back gate voltage. An increase of these voltages leads to an increase of the current magnitude and the saturation intensity. At room temperature the saturation is well described by the empirical formula $J_0\propto 1/(1+I/I_s)$ with the parameter $I_s \propto (1+\omega^2\tau^2)$. At low temperatures the behavior of the photocurrent becomes more complex and, in some cases, even exhibits a sign change with increasing radiation intensity. The complex behavior at low temperatures is attributed to the interplay of two microscopic mechanisms of the ratchet photocurrent formation: the Seebeck ratchet effect and the dynamic carrier-density redistribution mechanism. The top and bottom DGG structures exhibit a qualitatively similar overall functional behavior of the induced THz ratchet current. 
However, when comparing the magnitudes of the ratchet current at low power, it becomes evident that the bottom DGG devices exhibit significantly lower responsivity,
see Appendix~\ref{bottomDGG}. 
As a significant conclusion, we provide compelling evidence that the asymmetric graphene-based DGG structures, which have been recently proposed as high-speed, sensitive room temperature THz detectors, exhibit a linear photoresponse even at very high radiation intensities reaching the order of hundreds of kW\,cm$^{-2}$.

\section{Acknowledgments}
\label{acknow} 

The support of the Deutsche Forschungsgemeinschaft (DFG, German Research Foundation) Project No. Ga501/18, IRAP Programme of the Foundation for Polish Science (Grant No. MAB/2018/9, project CENTERA), and the Volkswagen Foundation (97738) is gratefully acknowledged. J.E. and R.H. acknowledge support of the Deutsche Forschungsgemeinschaft through SFB 1277 (Project Id 314695032, subProject No. A09), WE 2476/11-2 and GRK 1570. This project has received funding from the European Research Council (ERC) under the European Union’s Horizon 2020 research and innovation programme (grant agreement No 787515) (R.H.). K.W. and T.T. acknowledge support from the JSPS KAKENHI (Grant Numbers 20H00354, 21H05233 and 23H02052) and World Premier International Research Center Initiative (WPI), MEXT, Japan.


\appendix

\begin{figure}
	\centering
	\includegraphics[width=\linewidth]{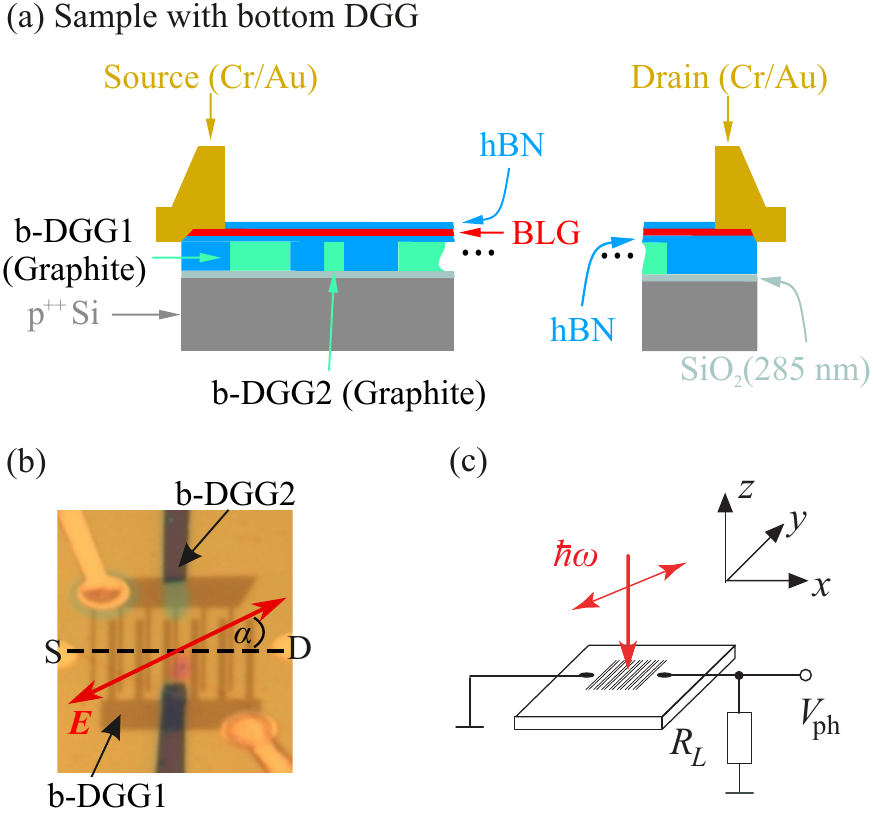}
	\caption{Panel (a) shows the cross section of the bottom DGG device. The inter-digitated dual-grating gate is made of graphite (green) separated by hBN (blue). 
		The hBN-encapsulated BLG (red) structure is stacked on top of the DGG. Panel (b) shows the optical mircophotograph of the structure. The azimuth angle $\alpha$ defines the orientation of the radiation electric field (red double arrow) in respect to S-D direction (horizontal dashed line). Panel (c) shows the measurement configuration. In the experiments with pulsed laser the voltage signal was picked up across a load resistance $R_{L} = 50~\Omega$, whereas for measurements with $cw$ laser the voltage drop was directly obtained across the sample resistance.}
	\label{Fig1samplebottomDGG}
\end{figure}

\begin{figure}
	\centering
	\includegraphics[width=\linewidth]{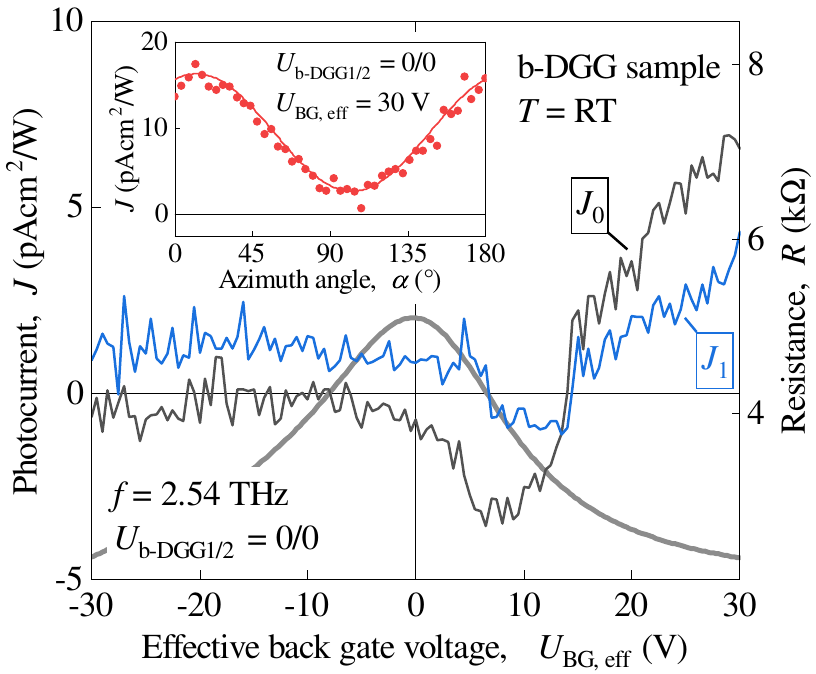}
	\caption{Dependencies of the photocurrent contributions $J_0$ and $J_1$ on the effective back gate voltage $U_{\text{\text{BG, eff}}}$ measured at low-power radiation with a frequency of \TeraHertz{2.54} at room temperature and both bottom gates at zero bias. The inset shows dependence of the photocurrent on the azimuth angle $\alpha$  at a back gate voltage of $U_{\text{BG,eff}}=\SI{30}{\volt}$, and zero bottom gate biases. It demonstrates that the ratchet current excited by linearly polarized radiation consists of two contributions $J_{0} = 9.2$~pA\,cm$^{2}$/W and $J_{1} = 6$~pA\,cm$^{2}$/W. Note that a possible contribution $J_2 \ll J_0, J_1$. The thick gray line depicts the corresponding sample resistance.
	}
	\label{FigRT1_alpha}
\end{figure}

\section{DCDR microscopic mechanism of the polarization independent and linear ratchet current}
\label{DCDR}

The dynamic carrier-density redistribution considers a space periodic profile of the carrier density induced by the THz radiation. The model of this effect has been discussed in Ref.~\cite{Ivchenko2011} in which it was focused on the origin of the linear ratchet effect. In general, however, the ratchet current has an additional polarization independent contribution $J_0$ which is of importance for the description of the nonmonotonic photocurrent intensity dependence. In the frame of this model the ac electric field $\bm E(x)\exp(-i\omega t)+c.c.$ with the near-field amplitude $\bm E(x)$ applied to the electronic system being in the space-periodic static potential $V(x)$ results in a space-periodic time-oscillating profile $\delta N(x,t)=\delta N_\omega(x)\exp(-i\omega t)+c.c.$ It satisfies the continuity equation
\begin{equation}
	-i\omega e\delta N_\omega(x) +\pdv{}{x}\delta\sigma_\omega(x)E_x(x)=0,
\end{equation}
where $\delta\sigma_\omega(x) = V(x)\partial \sigma_\omega/\partial \varepsilon_{\rm F}$ with $\sigma_\omega=\sigma_0/(1-i\omega \tau)$ being the complex conductivity.
This means that the following time-oscillating density profile is formed:
\begin{equation}
	\delta N(x,t) = {2\sigma_0\qty[\omega\tau\cos(\omega t) - \sin(\omega t)]\over e \omega \varepsilon_{\rm F} \qty[1+ (\omega\tau)^2]}\frac{dV(x)}{dx}E_x(x).
\end{equation}
This expression shows that due to retardation the density profile oscillates partially in phase with the electric field: $\delta N \propto \omega\tau\cos(\omega t)$.
Taking into account this density profile, the \textit{dc} electric current is given by
\begin{equation}
	j_x^{\rm DCDR} = \pdv{\sigma_0}{N} \overline{\delta N(x,t)E_x(x,t)},
\end{equation}
where the bar denotes averaging in both time and spatial period. With account for both valley and spin degeneracy of BLG this yields 
\begin{equation}
j_x^{\rm DCDR} = \Xi {2e^3\tau^3\over \pi \hbar^2 m\qty[1+ (\omega\tau)^2]} \abs{e_x}^2.
\end{equation}

Finally, we obtain the photocurrent containing both the polarization-dependent ($\propto P_L$) and independent parts, which have equal amplitudes:
\begin{equation}
j_x^{\rm DCDR} = \Xi {e^3\tau^3\over \pi \hbar^2 m\qty[1+ (\omega\tau)^2]} (1+P_L).
\end{equation}

\begin{figure}
	\centering
	\includegraphics[width=\linewidth]{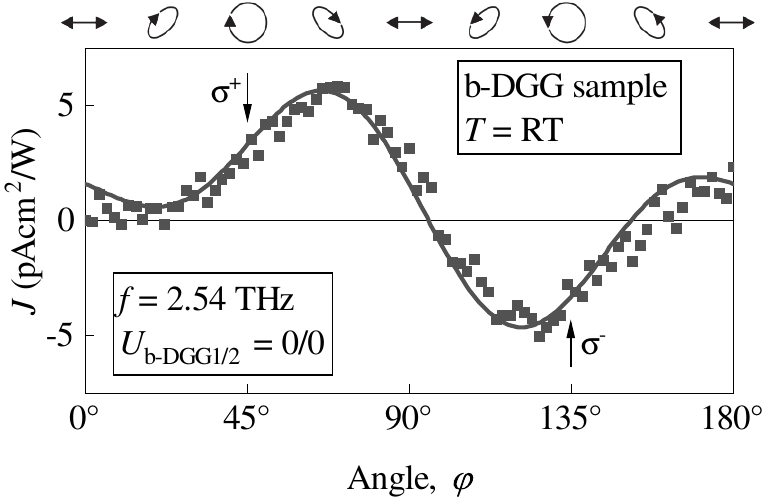}
	\caption{
		Polarization dependence of the photocurrent obtained by rotation of the lambda-quarter plate by angle $\varphi$. The data are obtained at low-power radiation of $cw$ laser operating at $f=2.54$~THz, room temperature and all gate voltages equal to zero. Upper arrows labeled as $\sigma^+$ and $\sigma^-$ indicate the photocurrent measured for right- and left-handed circularly polarized radiation, respectively. The solid curve shows the fit after Eq.~\ref{circular}. It demonstrates a large contribution of the circular ratchet current $J_{\rm circ} = 3.3$~pA\,cm$^{2}$/W.
		Double arrows, ellipses and circles on the top illustrate the state of polarization for several values of angle $\varphi$.
	}
	\label{FigRT1_circ}
\end{figure}

\section{Results on graphene devices with bottom asymmetric dual-grating gates}
\label{bottomDGG}

The main text is aimed to the ratchet effect in the top DGG structures, which have been studied extensively in relation to THz ratchet effect and THz detectors. Below we present the results of the study of the ratchet effects in novel design DGG structure that has been fabricated beneath bilayer graphene.
	
\subsection{Samples description}

Figure~\ref{Fig1samplebottomDGG}(a) shows the cross section of the structure. The bottom dual grating gate was fabricated out of a graphite (5-10 layers) located beneath the encapsulated BLG. After mechanical exfoliation of the graphite layer it was deposited onto a highly doped silicon \ce{Si}$^{++}$ with \NanoMeter{285} of \ce{SiO_2}. Subsequently, EBL was used to fabricate the pattern for the DGG within the graphite layer, followed by an oxygen plasma etching at low pressure, thereafter the process was finalized by a lift-off. As a next step, the hBN-encapsulated BLG was deposited directly on the interdigitated gate structures. The BLG flake has a size of 27~$\mu$m $\times$ 14~$\mu$m covered by 35~$\mu$m top and 40~$\mu$m bottom hBN. Lastly, the electrical contacts were fabricated by using the EBL technique, including a reactive ion etching system for etching of the hBN layers, and by metal deposition (\NanoMeter{0.2} of \ce{Cr}, and \NanoMeter{80} of \ce{Au}) on the BLG. An optical microphotograph of the device is shown in Fig.~\ref{Fig1samplebottomDGG}(b). The interdigitated periodic structure consists of two bottom gates: a wide (b-DGG1) and narrow (b-DGG2) gate with 1~$\mu$m / 0.5~$\mu$m width and 2~$\mu$m / 0.5~$\mu$m spacing parameters. Consequently, an asymmetric lateral lattice is obtained from six cells of recurring gate fingers and spacings in between with a characteristic period $L$. The bottom subgates are electrically isolated from each other, so that individual biasing of b-DGG1 and b-DGG2 was possible. Application of voltages $U_{\text{b-DGG1}}$ and $U_{\text{b-DGG2}}$ allows one to tune the lateral asymmetry parameter $\Xi$. The devices were excited by normal incident THz radiation and the photocurrent was measured from S and D contacts, see Figs.~\ref{Fig1samplebottomDGG}(b) and (c). 
Furthermore, as we show below, in the bottom DGG structures the lateral asymmetry parameter is also controlled by the unstructured back gate.


\begin{figure}
	\centering
	\includegraphics[width=\linewidth]{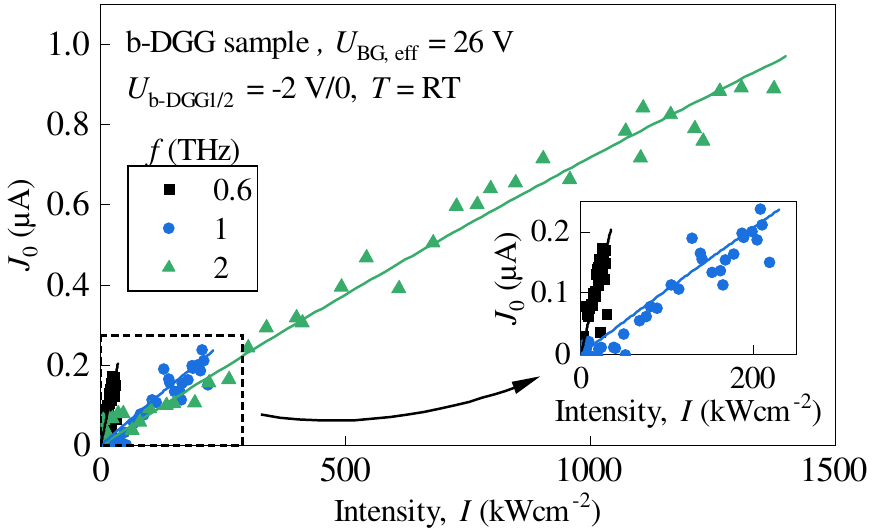}
	\caption{
		Dependencies of the photocurrent contribution $J_0$
		at a bottom gates voltage combination of $U_{\text{b-DGG1/2}}= \SI{-2}{\volt} / 0$. The data were obtained at several radiation frequencies and an effective back voltage of \SI{26}{\volt}. The inset presents a zoom-in of the data for $f = 0.6$~THz and $f = 1$~THz, at which the intensity of our laser system is limited to $\approx 60$~kW/cm$^2$ and 250~kW/cm$^2$, respectively. Solid curves are fits after Eq.~\ref{saturation_fit} with two fitting parameters. The low-power amplitude $A$ and the saturation intensity $I_s$, which are: for $f = 0.6$~THz - $A =6$~pA/(W/cm$^2$) and  $I_s= 1200$~kW/cm$^2$; for $f = 1$~THz - $A =1.08$~pA/(W/cm$^2$) and  $I_s= 4500$~kW/cm$^2$; and for $f = 2$~THz - $A =0.79$~pA/(W/cm$^2$) and  $I_s= 10000$~kW/cm$^2$.
	}
	\label{FigRT1}
\end{figure}

\subsection{Photocurrents at low-power excitation}

The photocurrent detected in the devices of this design shows all characteristic features of the ratchet effect and, in fact behaves similarly to that one excited in conventional top DGG devices described in the main text. The inset in Fig.~\ref{FigRT1_alpha} shows the dependence of the photocurrent on the orientation of the radiation electric field vector. The data are obtained for room temperature and low-power excitation with a radiation frequency of $f =$ 2.54~THz. Alike for the top DGG devices, the polarization dependence can be well fitted by Eq.~\ref{linear} with a negligible current contribution $J_2$. The comparison of the inset in Fig.~\ref{FigRT1_alpha} (b-DGG sample) with Fig.~\ref{lin_pol_dep}(a) (t-DGG sample) shows that both $J_0$ and $J_1$ detected in the bottom gate device are about one order of magnitude less than those in the top DGG structures. Even despite the fact that in the top DGG structures the back gate voltage was zero whereas the data in the bottom DGG were obtained for a high back gate voltage of $U_{\text{BG,eff}}=\SI{30}{\volt}$, see Fig.~\ref{FigRT1_alpha}. While being clearly detected and well described by Eq.~\ref{circular}, the circular ratchet photocurrents, see Fig.~\ref{FigRT1_circ}, is also about an order of magnitude less than that detected for the t-DGG devices, see Fig.~\ref{lin_pol_dep}. These observations demonstrate that, at room temperature, the bottom gate design is less favorable for THz radiation detection than the top DGG one. 
Despite the unfavorable reduction of the device responsivity we present the results obtained in this structure due to its novel design, which was not investigated and analyzed so far (all previous studies were aimed to the top DGG design). 

\begin{figure}
	\centering
	\includegraphics[width=\linewidth]{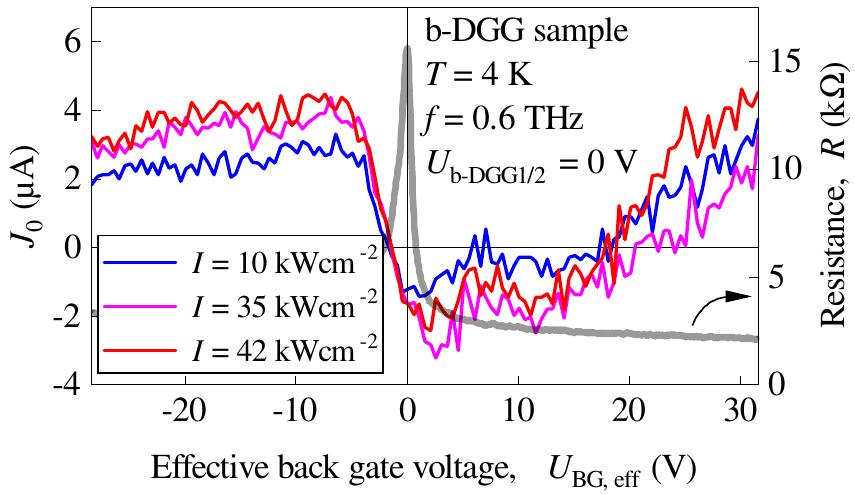}
	\caption{Dependencies of the photocurrent contributions $J_0$ on the effective back gate voltage $U_{\text{\text{BG, eff}}}$ measured at a radiation frequency of \TeraHertz{0.6}, different radiation intensities, and  a bottom gate voltage $U_{\text{b-DGG1/2}}= 0/ 0$. The thick gray line depicts the corresponding sample resistance.} 
	\label{Fig4new}
\end{figure}

\begin{figure}
	\centering
	\includegraphics[width=\linewidth]{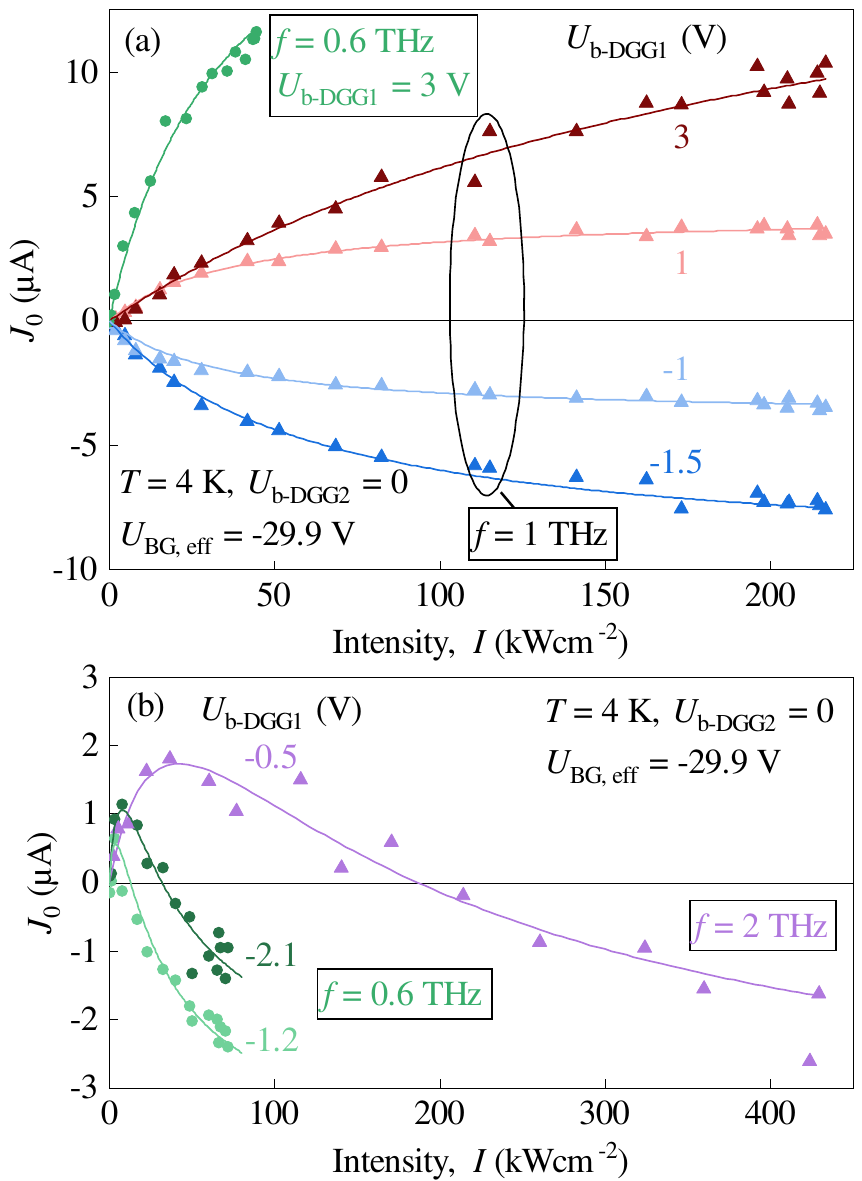}
	\caption{Dependencies of the photocurrent contribution $J_0$
		on the radiation intensity obtained for different frequencies an effective back voltage of \SI{-29.9}{\volt}, $U_{\text{b-DGG2}}= 0$ and different values of the bottom gate voltage $U_{\text{b-DGG1}}$ indicated by numbers close to each curve. Solid curves in panel (a) are fits after Eq.~\ref{saturation_fit}. The fitting parameters for $f = 0.6$~THz are 
		$A =0.6$~nA/(W/cm$^2$) and  $I_s= 35$~kW\,cm$^{-2}$ (3~V). Here and below the numbers in brackets are values of $U_{\text{b-DGG1}}$. The fitting parameters for $f = 1$~THz are $A =0.09$~nA/(W/cm$^2$) and  $I_s= 215$~kW\,cm$^{-2}$ (3~V); $A =0.115$~nA/(W/cm$^2$) and  $I_s= 37.9$~kW\,cm$^{-2}$ (1~V); $A =-0.115$~nA/(W/cm$^2$) and  $I_s= 33.7$~kW\,cm$^{-2}$ (-1~V); $A =-0.16$~nA/(W/cm$^2$) and  $I_s= 60$~kW\,cm$^{-2}$ (-1.5~V). Solid curves in panel (b) are fits after Eq.~\ref{signchange_fit} with fitting parameters: $A =0.94$~nA/(W/cm$^2$) and  $I_{s,A}= 70$~kW\,cm$^{-2}$, $B =-0.81$~nA/(W/cm$^2$) and  $I_{s,B}= 86$~kW\,cm$^{-2}$ (-1.5~V); $A =1.41$~nA/(W/cm$^2$) and  $I_{s,A}= 12.86$~kW\,cm$^{-2}$, $B =-1.01$~nA/(W/cm$^2$) and  $I_{s,B}= 21.2$~kW\,cm$^{-2}$ (-2.1~V); $A =0.75$~nA/(W/cm$^2$) and  $I_{s,A}= 5.4$~kW\,cm$^{-2}$, $B =-0.34$~nA/(W/cm$^2$) and $I_{s,B}= 24$~kW\,cm$^{-2}$ (-1.2~V). 
	}
	\label{Fig6}
\end{figure}

\begin{figure}[h!]
	\centering
	\includegraphics[width=\linewidth]{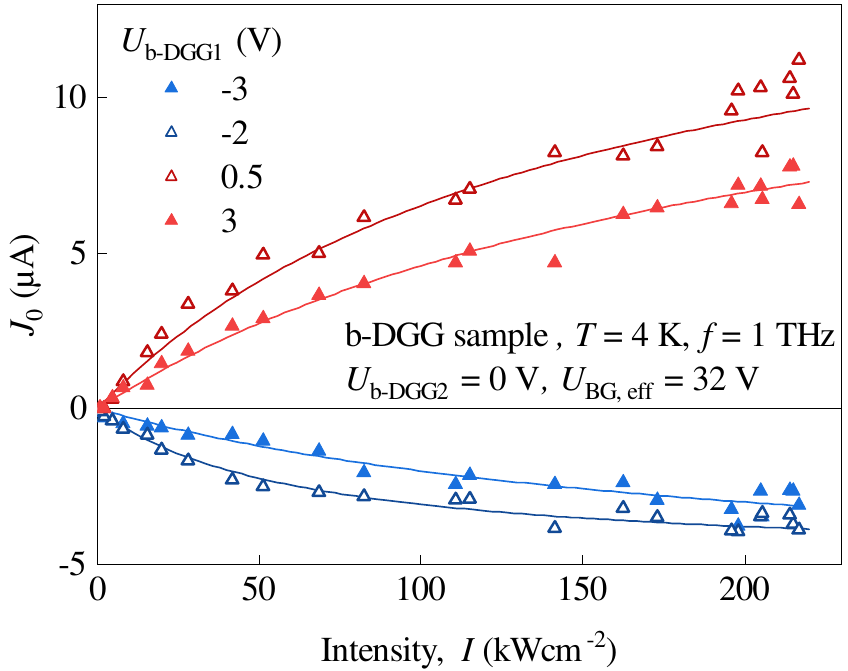}
	\caption{
		Dependencies of the photocurrent contribution $J_0$ on the radiation intensity. The data are obtained at a radiation frequency of 1~THz, effective back voltage of \SI{32}{\volt}, $U_{\text{b-DGG2}}= 0$, and different values of the bottom gate voltage  $U_{\text{b-DGG1}}$ indicted by numbers in the legend. Solid curves are fits after Eq.~\ref{saturation_fit} with fitting parameters: $A =-0.082$~nA/(W/cm$^2$) and  $I_s= 60$~kW\,cm$^{-2}$ (-2~V); $A =-0.11$~nA/(W/cm$^2$) and  $I_s= 20$~kW\,cm$^{-2}$ (-1.5~V); $A =0.11$~nA/(W/cm$^2$) and  $I_s= 146$~kW\,cm$^{-2}$ (0.5~V); $A =0.068$~nA/(W/cm$^2$) and  $I_s= 209$~kW\,cm$^{-2}$ (3~V). The numbers in brackets are values of $U_{\text{b-DGG1}}$. 
	}
	\label{Fig4}
\end{figure}

Figure~\ref{FigRT1_alpha} shows the back gate dependence of the photocurrent $J_0$. Strikingly, a variation of the effective back gate voltage results in a reversal of the photocurrent direction at rather high positive gate voltages $U_{\text{BG,eff}}=\SI{15}{\volt}$. This is in contrast to the results for the top gate structures, in which the sign inversion has been detected in the vicinity of the CNP, see Ref.~\cite{Olbrich2016,Moench2022} or, e.g., Fig.~\ref{Fig4_2}. This difference, however, can be explained by simple qualitative arguments. While the variation of the back gate homogeneously changes the carrier type and density in the t-DGG devices, the bottom gate potential applied to the BLG sheet in the b-DGG structure is periodically screened by the conductive graphite subgates. Consequently, a variation of the back gate voltage results in the changing of the lateral asymmetry parameter $\Xi$ and may even result in the formation of lateral $p-n$ junctions. The interplay between the density modulation and the changes of the lateral parameter asymmetry may result in a complex back gate dependence, e.g., the expected change of the photocurrent sign in the vicinity of CNP may be compensated by simultaneous changes of the magnitude and sign of the parameter $\Xi$.


\begin{figure}
	\centering
	\includegraphics[width=\linewidth]{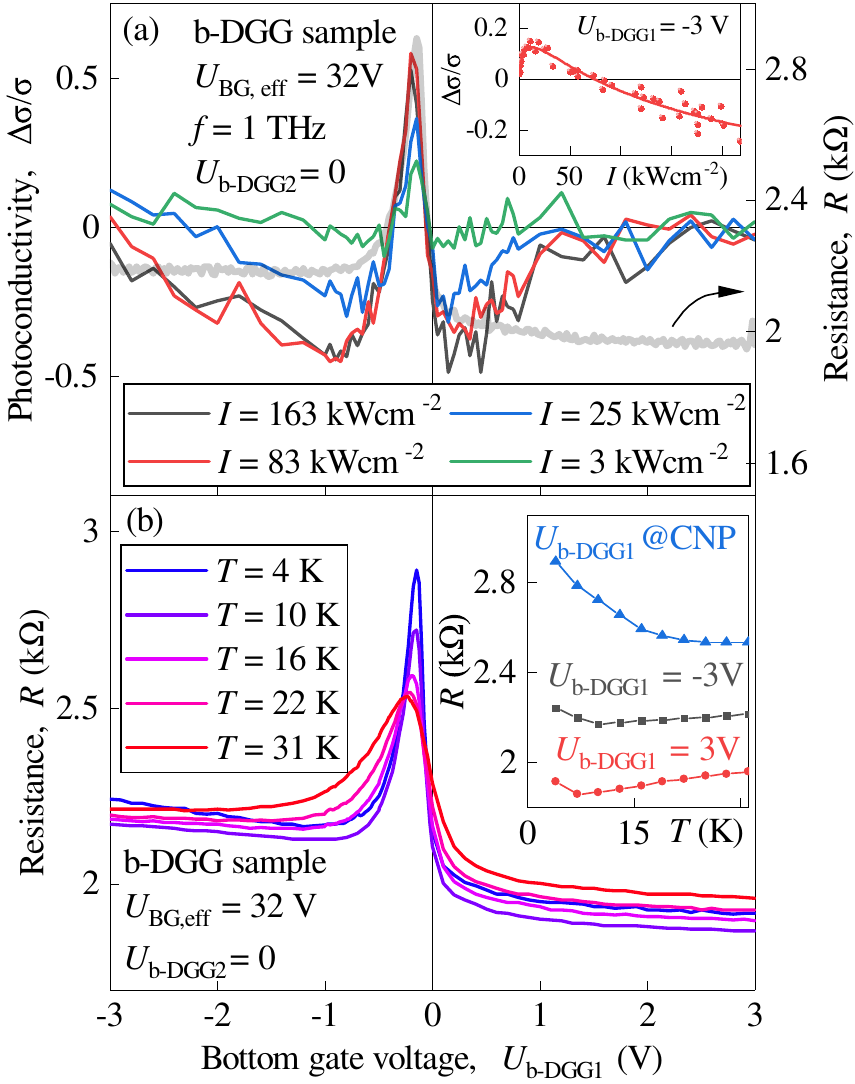}
	\caption{
		Panel (a): Dependencies of the photoconductivity on the bottom gate voltage $U_{\text{\rm b-DGG1}}$ ($U_{\text{\rm b-DGG2}}=0$) at different radiation intensities and an effective back gate voltage of \SI{32}{\volt}. The inset shows the intensity dependence of the photoconductivity measured for a bottom gate voltage of $U_{\text{\rm b-DGG1}}=\SI{-3}{\volt}$. Solid line is a guide for the eye. Panel (b): Dependencies of the sample resistance on the bottom gate voltage obtained at different temperatures. The inset shows temperature dependencies of the resistance measured at different bottom gate voltages $U_{\text{\rm b-DGG1}}$.}
	\label{Fig7}
\end{figure}

\subsection{High-power induced ratchet currents}

For high intensities the qualitative behavior of the bottom gate device is similar to that in top DGG structures. Figure~\ref{FigRT1} shows the intensity dependence of the photocurrent $J_0$ measured at room temperature at several radiation frequencies. While the data can be well described by Eq.~\ref{saturation_fit}, the saturation intensities are drastically higher than that detected for top DGG structures and the current behaves almost linear at the whole range of radiation intensities. 

At low temperature ($T=4$~K) the nonlinearity is clearly observed and the overall behavior is well described by Eq.~\ref{saturation_fit}, see Figs.~\ref{Fig6}(a) and \ref{Fig4}. This type of nonlinearity is detected for almost the whole range of the back gate voltages, see Fig.~\ref{Fig4new}, showing the dependence of $J_0$ on $U_{\text{BG,eff}}$ measured for different radiation intensities while keeping both bottom DGGs at zero bias. Figure~\ref{Fig4new} demonstrates that the photocurrent reverses its sign in the vicinity of the CNP as well as at high back gate voltages, i.e., behaves qualitatively similar to $J_0$ detected at room temperature, see Fig.~\ref{FigRT1_alpha}. Note that now the ratchet current amplitudes remain lower, but become comparable to that for top DGG. Also the saturation intensities at $T = 4$~K, which are much lower than that detected at room temperature become comparable to that in the top DGG device. These findings are illustrated in Figs.~\ref{Fig6}(a) and \ref{Fig4} for data obtained at $U_{\text{BG,eff}}=\SI{-29.9}{\volt}$ and $\SI{32}{\volt}$, respectively. Alike in the top DGG device for a certain gate sequences we observed that the photocurrent changes its sign with increasing radiation intensity, see Fig.~\ref{Fig6}(b), indicating that two mechanisms contribute to the photocurrent formation. 

\subsection{THz photoconductivity}

To investigate the electron gas heating we studied the THz-radiation induced photoconductivity measured in bottom DGG devices. Figure~\ref{Fig7} shows the bottom gate voltage dependence of the normalized photoconductivity $\Delta\sigma/\sigma$ measured at different intensity levels. In the vicinity of the CNP the photoconductivity is positive and is mostly caused by the interband optical transitions resulting in the generation of electron-hole pairs, 
Far from the CNP, this process vanishes, because the Fermi energy increases and becomes higher than the radiation photon energy, and, consequently possible final states of the direct optical transitions become occupied. Therefore, at high gate voltages the THz-photoconductivity is caused by the change of the sample's mobility due to the electron gas heating ($\mu-$photoconductivity or bolometric effect).	
In this bottom gate voltage range, the photoconductivity is negative at high intensities. However, at low intensities the photoconductivity is negative at low $U_{\text{b-DGG1}}$, but changes its sign at gate voltages $\abs{U_{\text{b-DGG1}}}$ between 1 and 3~V and becomes positive at higher gate voltages, see blue and green curves in Fig.~\ref{Fig7}(a). This change of sign is also seen in the intensity dependence of the photoconductivity, which is displayed in the inset in Fig.~\ref{Fig7}(a). The observed sign inversion of the photoconductive response reflects the change of the scattering mechanisms due to the variation of the bottom gate voltage (inversion of sign at low intensities and $U_{\text{b-DGG1}}$ between -1 and -3~V), or even the radiation intensity [inset in Fig.~\ref{Fig7}(a)]. To explore this feature we measured the bottom gate dependence of the \textit{dc} resistance for different temperatures, see Fig.~\ref{Fig7}(b). In addition, the inset shows temperature dependencies of the resistance extracted from the traces in Fig.~\ref{Fig7}(b) for three bottom gate voltages. This plot reveals that while at the CNP the sample resistance gradually decreases with increasing the temperature, at high negative/positive bottom gate voltages it behaves nonmonotonic: it first decreases with raising $T$, approaches minimum, and then starts to increase. This behavior indicates a temperature dependent change of sign of the first derivative ${\partial{R}}/{\partial{T}}$ (${\partial{\sigma}}/{\partial{T}}$). In photoconductive measurements this change of behavior results in a transition from the positive to the negative photoconductivity. Such a transition is clearly seen in the inset in Fig.~\ref{Fig7}(a) demonstrating that an increase of the radiation intensity results in the change of the sign of $\Delta\sigma/\sigma$.

Comparing the intensity dependencies of the photoconductive signal and the ratchet effect, we see that alike in the top DGG structures they may behave qualitatively  different.  For instance,  while the photocurrent saturates and remains without a sign inversion, the photoconductivity changes its sign, see  Fig.~\ref{Fig4} and  inset in Fig.~\ref{Fig7}(a), respectively. This observation indicates that also in the bottom DGG the interplay of the Seebeck  and DCDR mechanisms of the ratchet effects play an important role, see Sec.~\ref{high_power_top_gate}.


\subsection{Summary of the result on the bottom DGG devices}

Our results demonstrate that the general behavior of the ratchet effect generated in the bottom DGG devices behaves similar to that in the conventional top DGG device. There is only a qualitative difference in the action of the back gate, which modifies the lateral asymmetry parameter $\Xi$ in the bottom DGG structure.
Our studies indicate that the responsivity of devices with bottom DGG structures fabricated of graphite gates was significantly lower compared to those with top DGG structures produced with highly conductive gold film gates.
The difference in the conductivity of the gate materials may be responsible for the lower response detected in the bottom DGG devices.


\bibliography{all_lib}

\end{document}